\documentclass[fleqn,usenatbib]{mnras}

\usepackage{newtxtext,newtxmath}

\usepackage[T1]{fontenc}

\DeclareRobustCommand{\VAN}[3]{#2}
\let\VANthebibliography\thebibliography
\def\thebibliography{\DeclareRobustCommand{\VAN}[3]{##3}\VANthebibliography}

\usepackage{graphicx}	
\usepackage{amsmath}	
\usepackage{multicol}
\usepackage{deluxetable}
\usepackage{subcaption}

\title[Outflow of J2357-0048]{VLT/UVES Observation of the SDSS J2357-0048 Outflow}

\author[D. Byun et al.]{
Doyee Byun,$^{1}$\thanks{E-mail: dbyun@vt.edu (DB)}
Nahum Arav,$^{1}$
Patrick B. Hall$^{2}$
\\
$^{1}$Department of Physics, Virginia Tech, Blacksburg, VA 24061, USA\\
$^{2}$Department of Physics and Astronomy, York University, Toronto, ON
M3J 1P3, Canada
}

\date{Accepted XXX. Received YYY; in original form ZZZ}

\pubyear{2022}

\begin{document}
\label{firstpage}
\pagerange{\pageref{firstpage}--\pageref{lastpage}}
\maketitle

\begin{abstract}
We found a broad absorption line (BAL) outflow in the VLT/UVES spectrum of the quasar SDSS J235702.54-004824.0, in which we identified four subcomponents. We measured the column densities of the ions in one of the subcomponents ($v=-1600\text{ km s}^{-1})$, which include \ion{O}{I} and \ion{Fe}{II}. We found the kinetic luminosity of this component to be at most $\sim2.4\%$ of the quasar's Eddington luminosity. This is near the amount required to contribute to AGN feedback. We also examined the time-variability of a \ion{C}{IV} mini-BAL found at $v=-8700\text{ km s}^{-1}$, which shows a shallower and narrower absorption feature attached to it in previous SDSS observations from 2000 and 2001, but not in the spectra from 2005 and onwards.
\end{abstract}




\section{Introduction}

Quasar outflows are often seen as absorption troughs blueshifted relative to the quasar rest frame in $\lesssim40\%$ of quasar spectra \citep{2003AJ....125.1784H,2008ApJ...672..108D,2008MNRAS.386.1426K}, often invoked as likely producers of AGN feedback \citep[e.g.,][]{1998A&A...331L...1S,2004ApJ...608...62S,2018ApJ...857..121Y,2021ApJ...919..122V,2022arXiv220206227H}. Theoretical estimates require these outflows to have a kinetic luminosity ($\dot{E}_k$) of $\sim0.5\%$ \citep{2010MNRAS.401....7H} or $\sim5\%$ \citep{2004ApJ...608...62S} of the quasar Eddington luminosity ($L_{Edd}$) to significantly contribute to AGN feedback. We have reason to interpret the luminosity mentioned by \citet{2010MNRAS.401....7H} and \citet{2004ApJ...608...62S} to be $L_{Edd}$ and not the bolometric luminosity ($L_{Bol}$), as explained in Section 4 of \citet{2020MNRAS.499.1522M}. Outflows satisfying these conditions have been found in previous studies \citep[e.g.,][]{2009ApJ...706..525M,2013MNRAS.436.3286A,2020ApJS..247...37A,2015MNRAS.450.1085C,2018ApJ...866....7L,2019ApJ...876..105X,2020ApJS..247...38X,2020ApJS..247...42X,2020ApJS..247...39M,2020ApJS..249...15M,2022Byun}.\par
An outflow's kinetic luminosity depends on its distance from the quasar ($R$). One way to estimate this distance is by measuring the ionization parameter ($U_H$) and electron number density ($n_e$) \citep[e.g.,][]{2012ApJ...758...69B}. Several studies in the past have employed this method to find the distance of outflow systems \citep[e.g.,][]{2001ApJ...548..609D,2002ApJ...567...58D,2001ApJ...550..142H,2005ApJ...631..741G,2012ApJ...758...69B,2018ApJ...858...39X,2020ApJS..247...37A,2020MNRAS.499.1522M,2020ApJS..247...39M,2022Byun}. The value of $n_e$ can be determined by finding the ratio between the column densities of energy states of the same ion \citep[e.g.,][]{2018ApJ...857...60A}.\par
This paper presents the determination of the $R$ and $n_e$ values, along with $\dot{E}_k$, of an outflow system found in the VLT/UVES spectrum of SDSS J235702.54-004824.0 (hereafter J2357-0048). The analysis in this paper has been conducted with data from the VLT/UVES Spectral Quasar Absorption Database (SQUAD) published by \citet{Murphy2019}, from which data has been retrieved for a similar analysis of quasar outflows \citep{2022Byun}.\par
The UVES data of J2357-0048 is from program 075.B-0190(A), and has been added to SQUAD by \citet{Murphy2019} and examined by \citet{2021ApJ...907...84C} for mini-BAL systems. \citet{2021ApJ...907...84C} report the presence of four absorption systems, which we independently identified as four sub-components of a BAL outflow. We found the lowest velocity system suitable for our analysis, as it shows troughs of excited states of \ion{Fe}{ii}, \ion{O}{i}, and \ion{Si}{ii}. We also found a high velocity \ion{C}{iv} mini-BAL, of which we report the time variability in comparison with SDSS spectra from different epochs.\par
This paper is structured as follows. Section \ref{sec:data} discusses the observation and data acquisition of J2357-0048. Section 
\ref{sec:analysis} shows the ionic column density measurements and the $n_e$ and $U_H$. In Section \ref{sec:results}, we present the analysis results and the energetics parameters. We also show the time variability analysis of the high velocity mini-BAL. In Section \ref{sec:discussion}, we discuss the results and other features of the quasar's spectrum, and Section \ref{sec:conclusion} summarizes and concludes the paper. For this analysis, we adopted a cosmology of $h=0.696$, $\Omega_m = 0.286$, and $\Omega_\Lambda = 0.714$ \citep{Bennett_2014}, and used the Python astronomy package Astropy \citep{astropy:2013,astropy:2018} for cosmological calculations.

\section{Observation and Data Acquisition} \label{sec:data}
The quasar J2357-0048 (J2000: RA=23:57:02.54, DEC=-00:48:24.0; $z=2.998$) was observed with the VLT/UVES on September 5, 2005 as part of the program 075.B-0190(A), with resolution $R\simeq 40,000$ and wavelength coverage from 3621 to 10429 \text{\AA \ } \citep{Murphy2019}.\par
\citet{Murphy2019} report a systemic redshift of $z=2.998$, also reporting values of $z=2.998$, $z=3.012$, $z=3.005$ from the SDSS, NED, and SIMBAD databases respectively. We found a discrepancy between these values and the redshifts that we found from the databases independently. From the SDSS data, we found redshifts ranging from $z=3.006$ \citep[ninth data release][]{2012ApJS..203...21A} to $z=3.062$ \citep[13th data release][]{2017ApJS..233...25A}. NED reports $z=3.063$ citing the 13th SDSS data release, while in SIMBAD we found $z=3.006$ from the ninth data release. While the literature reports a wide range of redshifts, we found that the systemic redshift $z=2.998$ reported by \citet{Murphy2019} is consistent with the \ion{C}{iv} emission redshift $z=2.97$, blueshifted relative to the systemic redshift, that we found based on the emission from the SDSS spectrum. We have also found a redshift of $z=2.99$ with the \ion{C}{iii}] emission complex, consistent with the adopted redshift of $z=2.998$.\par
\begin{figure*}
    \centering
    \includegraphics[width=\linewidth]{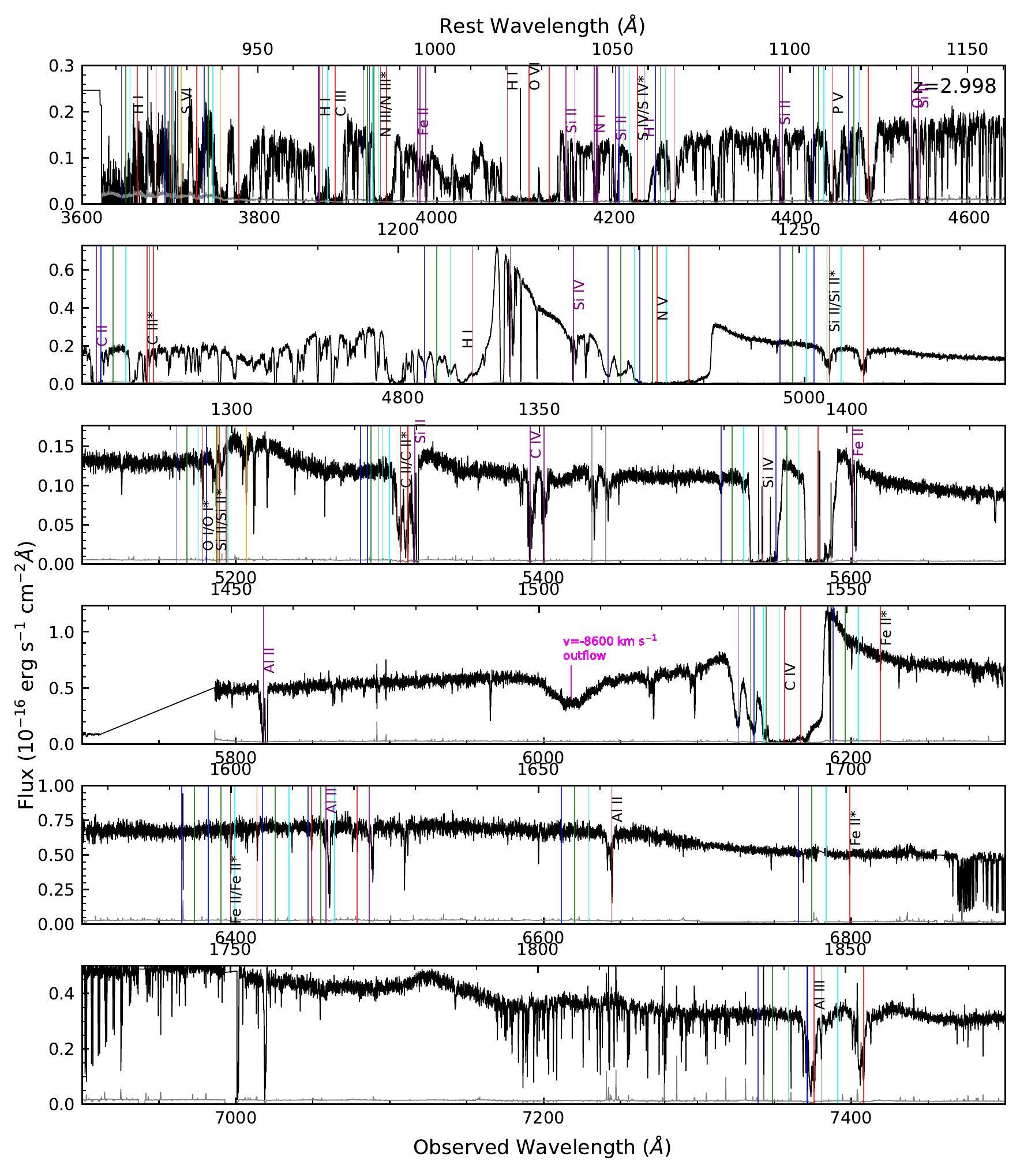}
    \caption{Normalized flux of J2357-0048 multiplied by the emission model by \citet{Murphy2019}, based on the SQUAD data set. The flux has been scaled to match the BOSS spectrum from the epoch of MJD=52203 (Oct. 21, 2001) at observed wavelength $\lambda = 6500$ \AA. The black curve represents the flux, and the gray shows the error in flux. The red, cyan, green, and blue vertical lines mark absorption troughs of outflow systems S1, S2, S3, and S4 respectively, while the S5 \ion{C}{iv} mini-BAL is labeled in magenta. Absorption features of the DLA system are marked in dark purple, and other intervening features are marked in grey. The \ion{Si}{ii} $\lambda\lambda$1304 multiplet of S1 is marked in orange to avoid confusion with the \ion{O}{i} multiplet troughs. The \ion{S}{vi} doublet of S1 is also marked in orange to avoid confusion with the \ion{H}{i} lines. We also marked where the absorption of the \ion{C}{iii}* multiplet near rest wavelength 1175\AA$\text{ would be found. Note that we cannot positively identify \ion{C}{iii}* due to the Lyman-$\alpha$ contamination.}$}
    \label{fig:fluxplot}
\end{figure*}

\citet{Murphy2019} reduced and normalized the UVES data by its continuum and emission as part of the SQUAD database. In Fig.~\ref{fig:fluxplot}, we show the normalized spectrum multiplied by the emission model, scaled to match the continuum level at observed wavelength $\lambda=6500$ \text{\AA \ } of the SDSS spectrum at the epoch of MJD=25503. \citet{2013A&A...556A.140Z} report the detection of a damped Ly $\alpha$ (DLA) system at $z=2.479$, and \citet{2021ApJ...907...84C} have identified four outflow absorption systems, which we identify here as four subcomponents of a BAL outflow (S1 at $v=-1600\text{ km s}^{-1}$, S2 at $v=-2300\text{ km s}^{-1}$, S3 at $v=-2700\text{ km s}^{-1}$, and S4 at $v=-3100\text{ km s}^{-1}$). We also identified a high velocity \ion{C}{iv} mini-BAL at $z=2.8849$, which we call S5 in this paper.\par
\begin{figure*}
    \centering
    \begin{multicols}{3}
    \subcaptionbox{\ion{H}{i}\label{fig:HI}}{\includegraphics[width=0.33\textwidth]{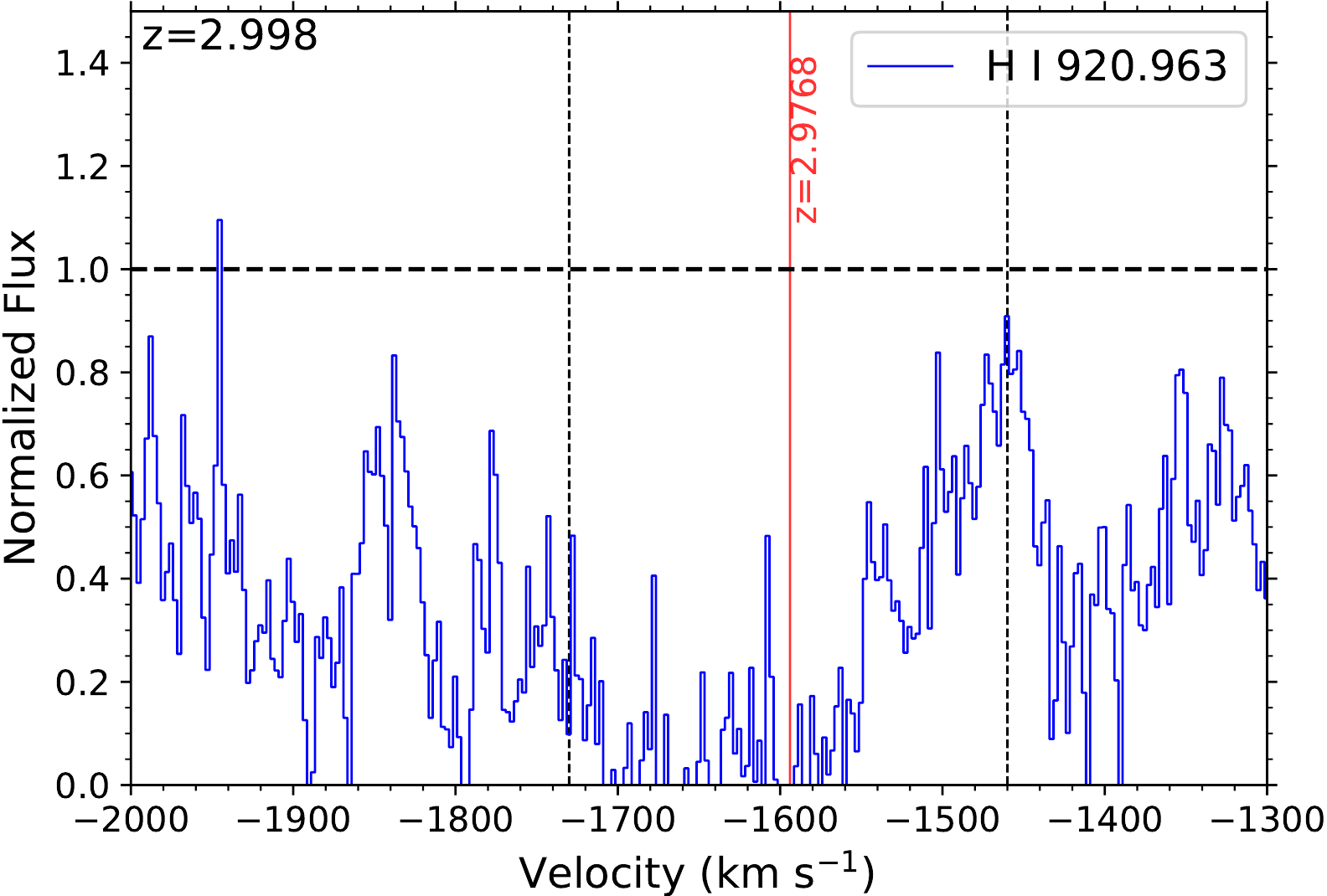}}\par
    \subcaptionbox{\ion{N}{v}\label{fig:NV}}{\includegraphics[width=0.33\textwidth]{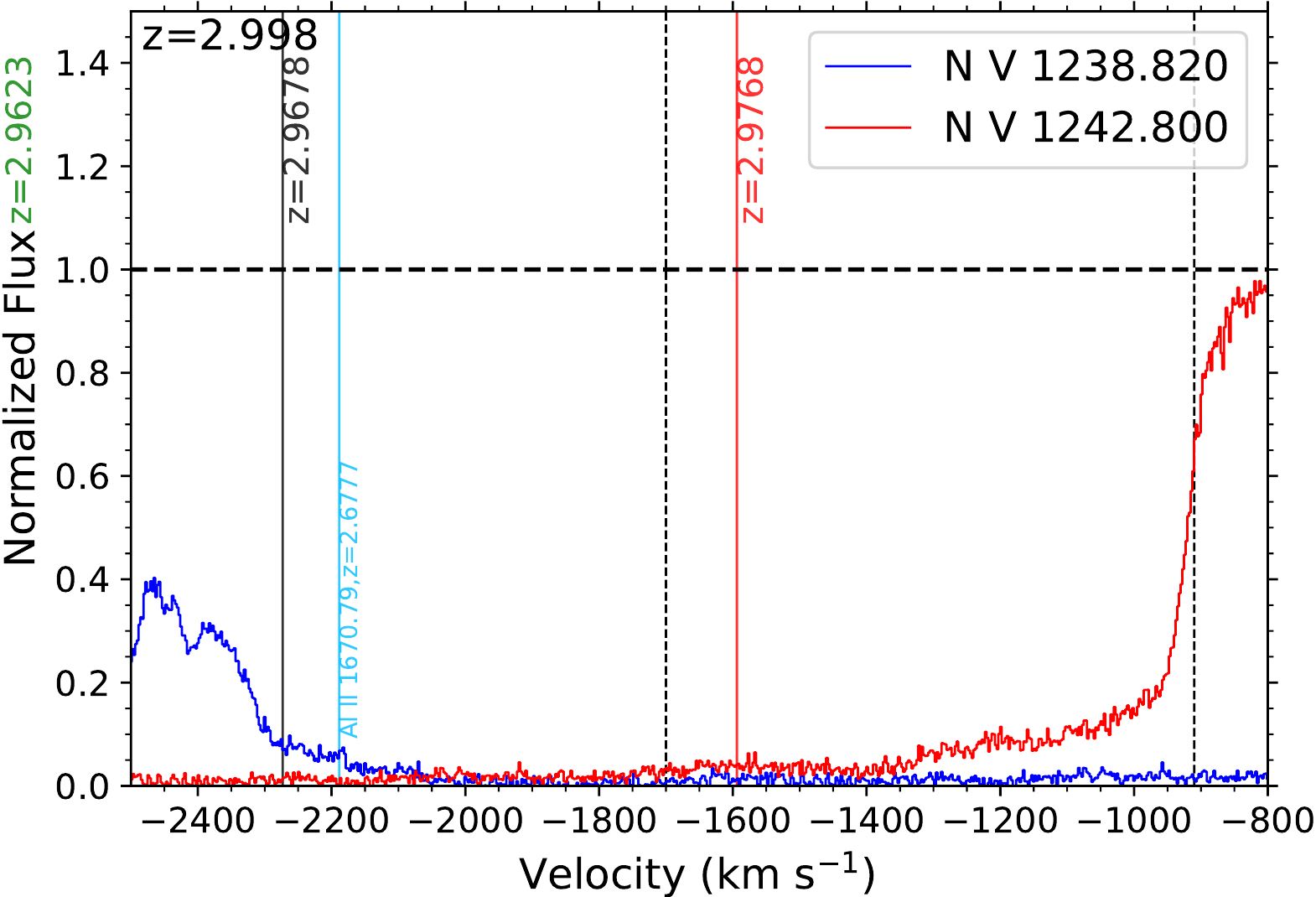}}\par
    \subcaptionbox{\ion{P}{v}\label{fig:PV}}{\includegraphics[width=0.33\textwidth]{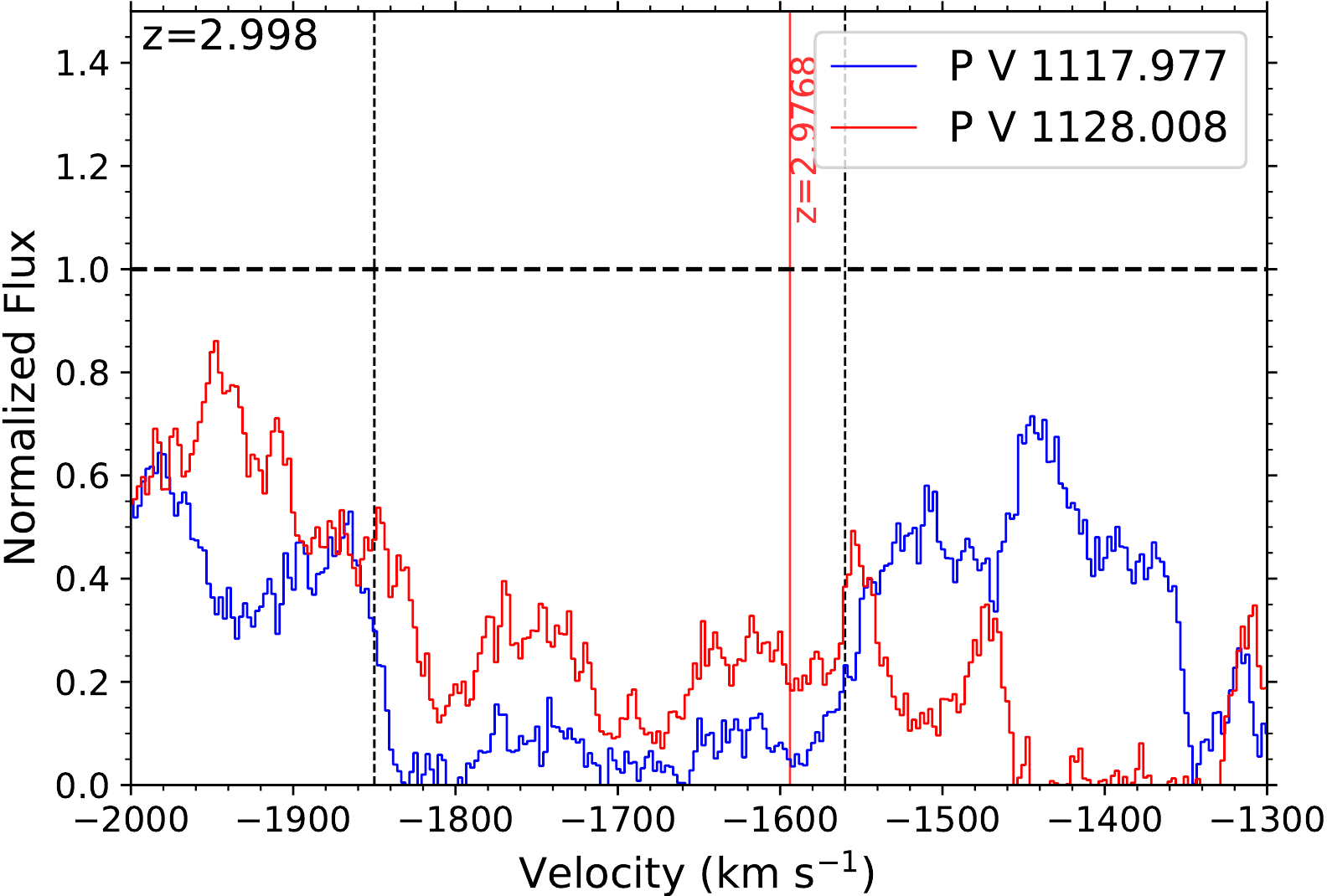}}\par
    \end{multicols}
    \begin{multicols}{3}
    \subcaptionbox{\ion{S}{vi}\label{fig:SVI}}{\includegraphics[width=0.33\textwidth]{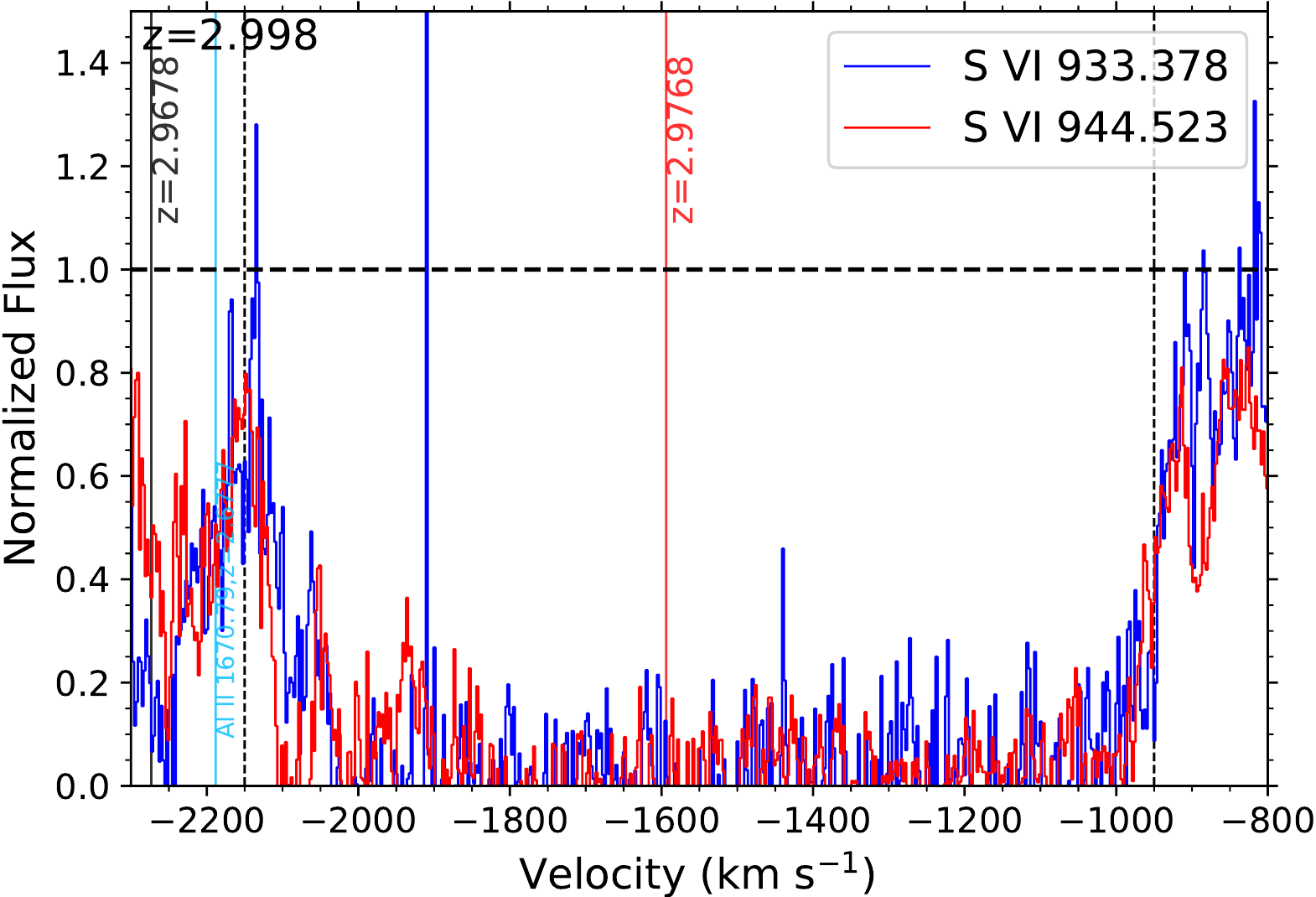}}\par
    \subcaptionbox{\ion{C}{iii}\label{fig:CIII}}{\includegraphics[width=0.33\textwidth]{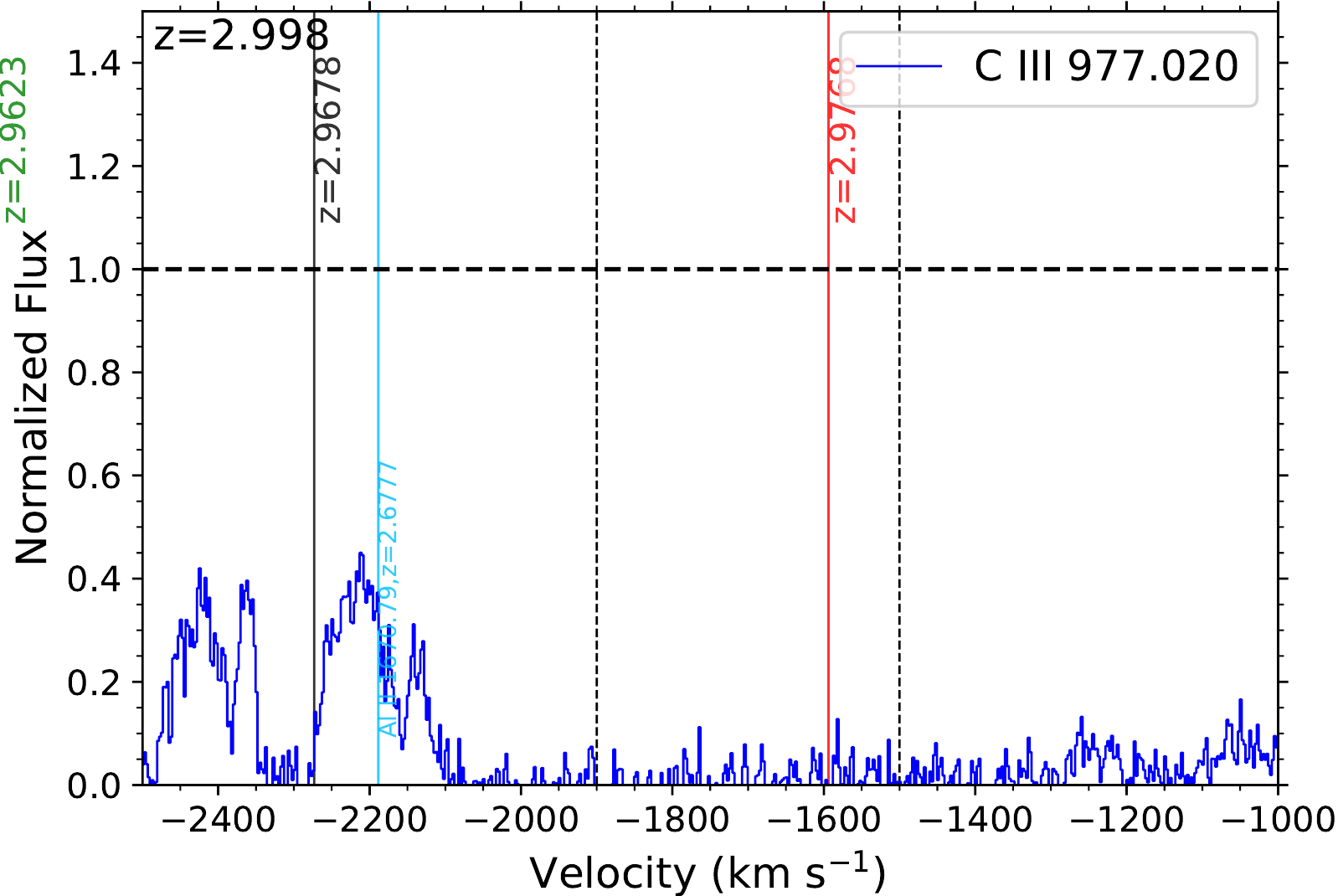}}\par
    \subcaptionbox{\ion{C}{iv}\label{fig:CIV}}{\includegraphics[width=0.33\textwidth]{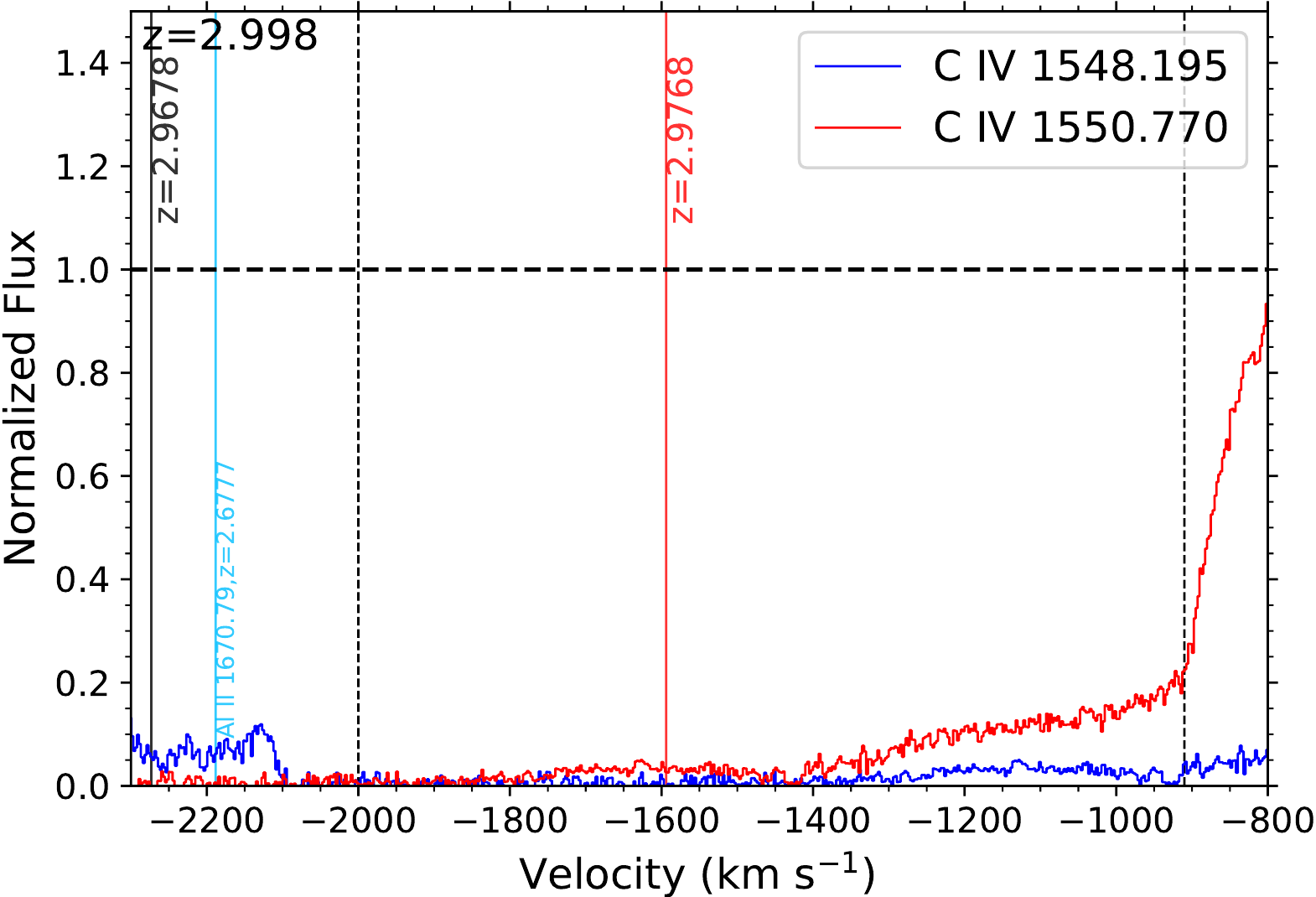}}\par
    \end{multicols}
    \begin{multicols}{3}
    \subcaptionbox{\ion{Si}{iv}\label{fig:SiIV}}{\includegraphics[width=0.33\textwidth]{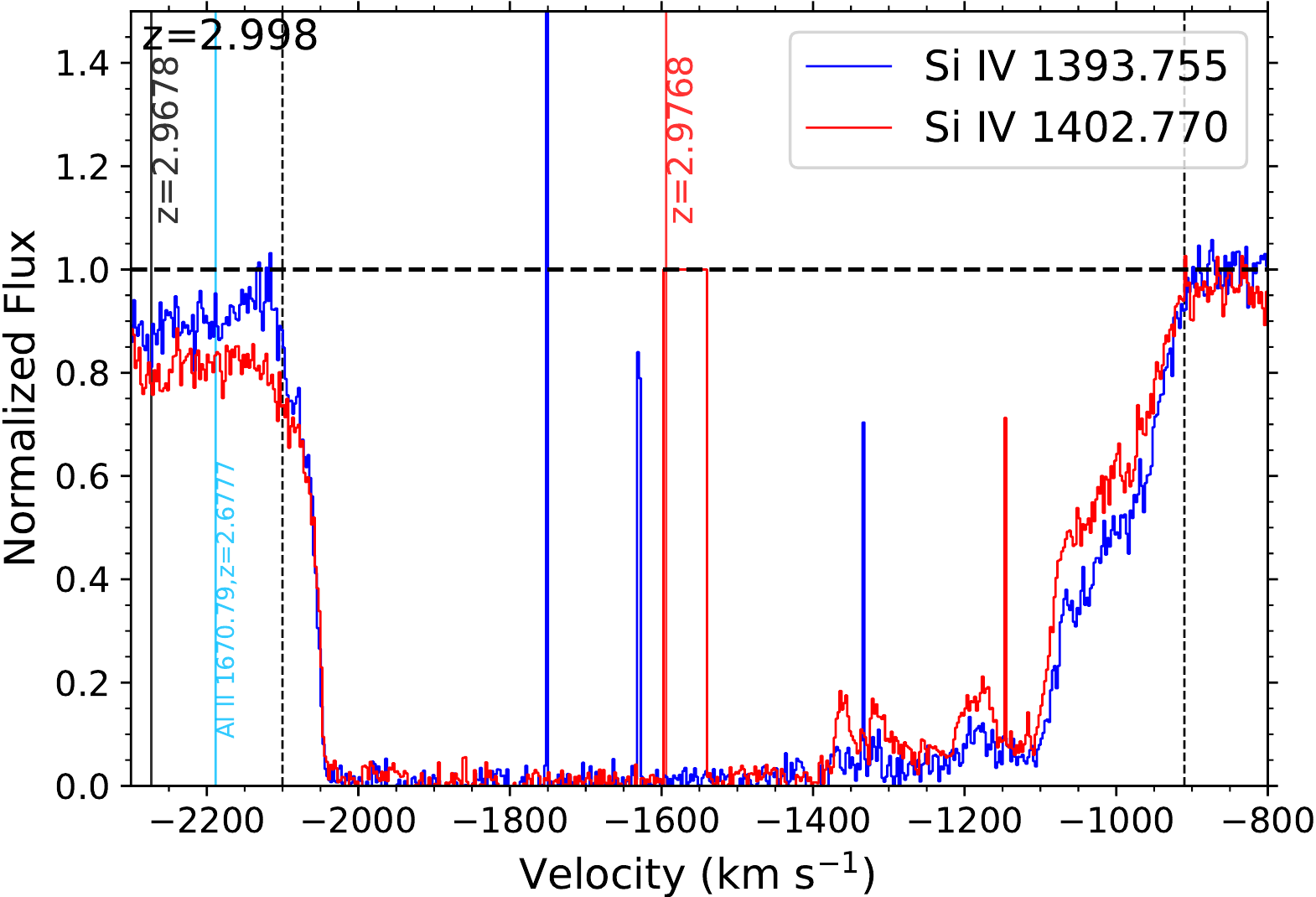}}\par
    \subcaptionbox{\ion{Al}{ii}\label{fig:AlII}}{\includegraphics[width=0.33\textwidth]{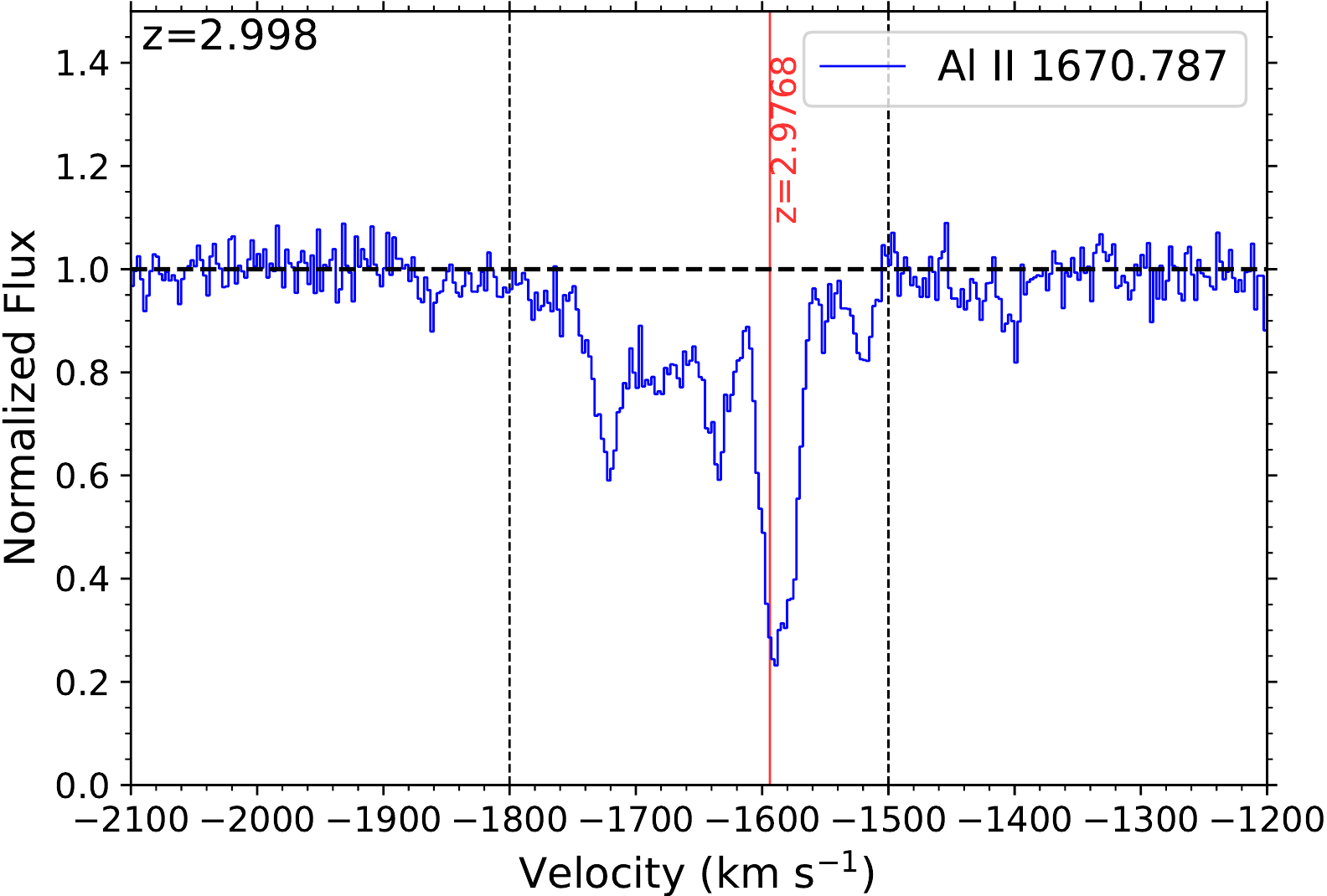}}\par
    \subcaptionbox{\ion{Al}{iii}\label{fig:AlIII}}{\includegraphics[width=0.33\textwidth]{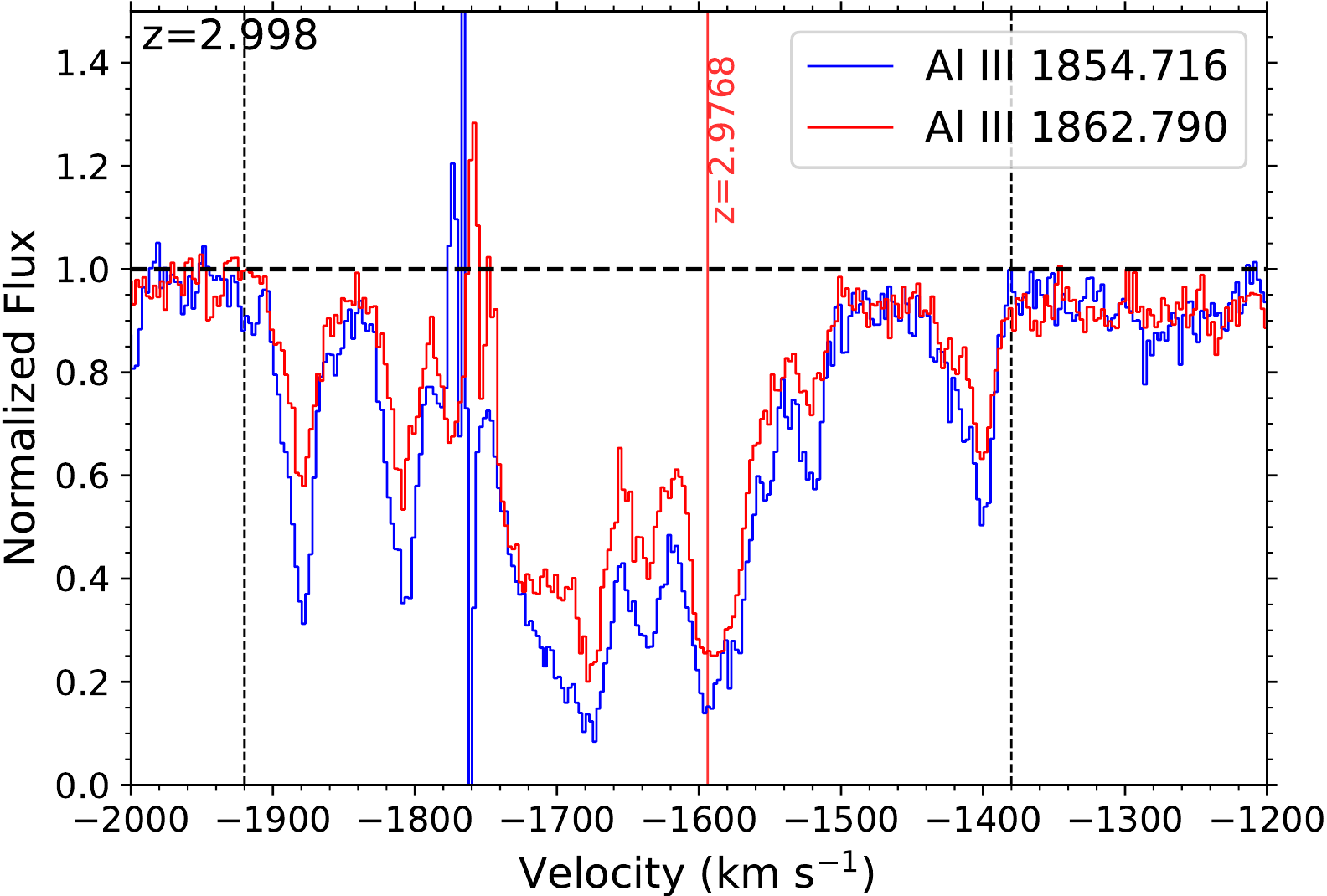}}\par
    \end{multicols}
\caption{Normalized spectra plotted in velocity space for ions in S1. The velocities of S1 ($z=2.9768$) and S2 ($z=2.9678$) are marked with red and black vertical lines respectively. The dotted vertical lines show the integration ranges used for column density calculations, while the horizontal dashed line marks the continuum level. The light blue vertical lines mark the intervening absorption systems that contaminate the blue spectra.}
\label{fig:vcut}
\end{figure*}
\begin{figure*}
    \centering
    \begin{multicols}{3}
    \subcaptionbox{\ion{N}{iii}\label{fig:NIII}}{\includegraphics[width=0.33\textwidth]{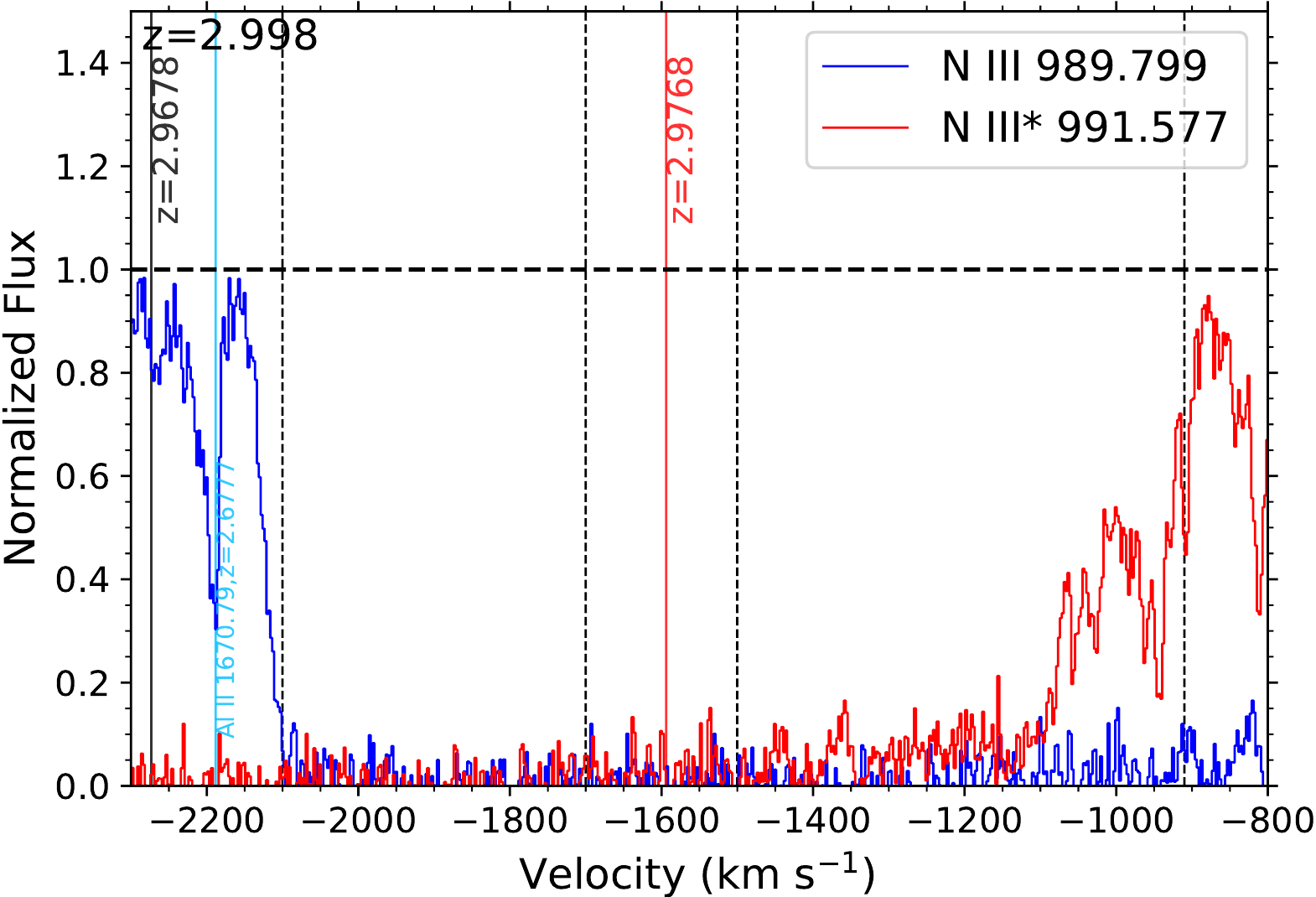}}\par
    \subcaptionbox{\ion{O}{i}\label{fig:OI}}{\includegraphics[width=0.33\textwidth]{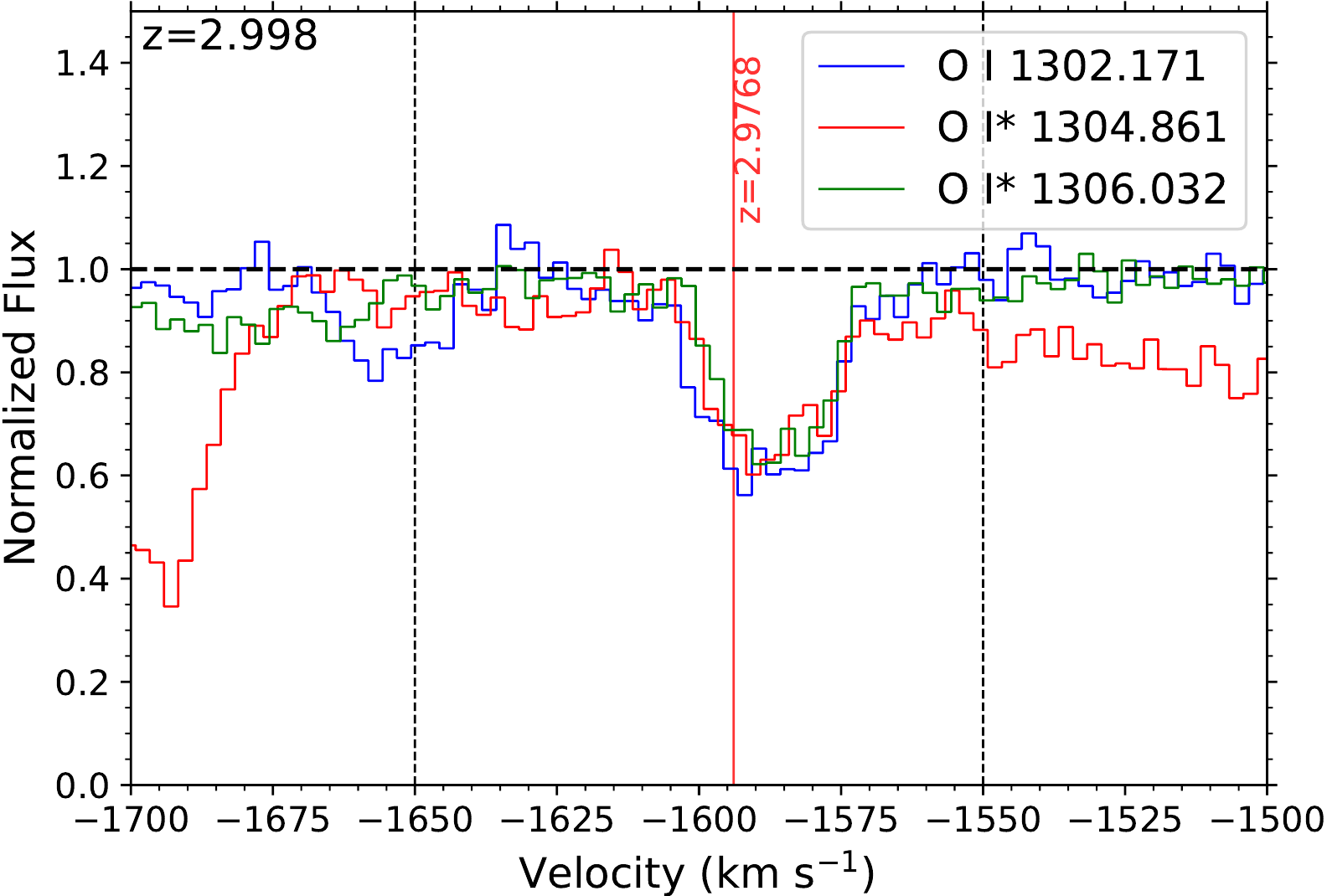}}\par
    \subcaptionbox{\ion{S}{iv}\label{fig:SIV}}{\includegraphics[width=0.33\textwidth]{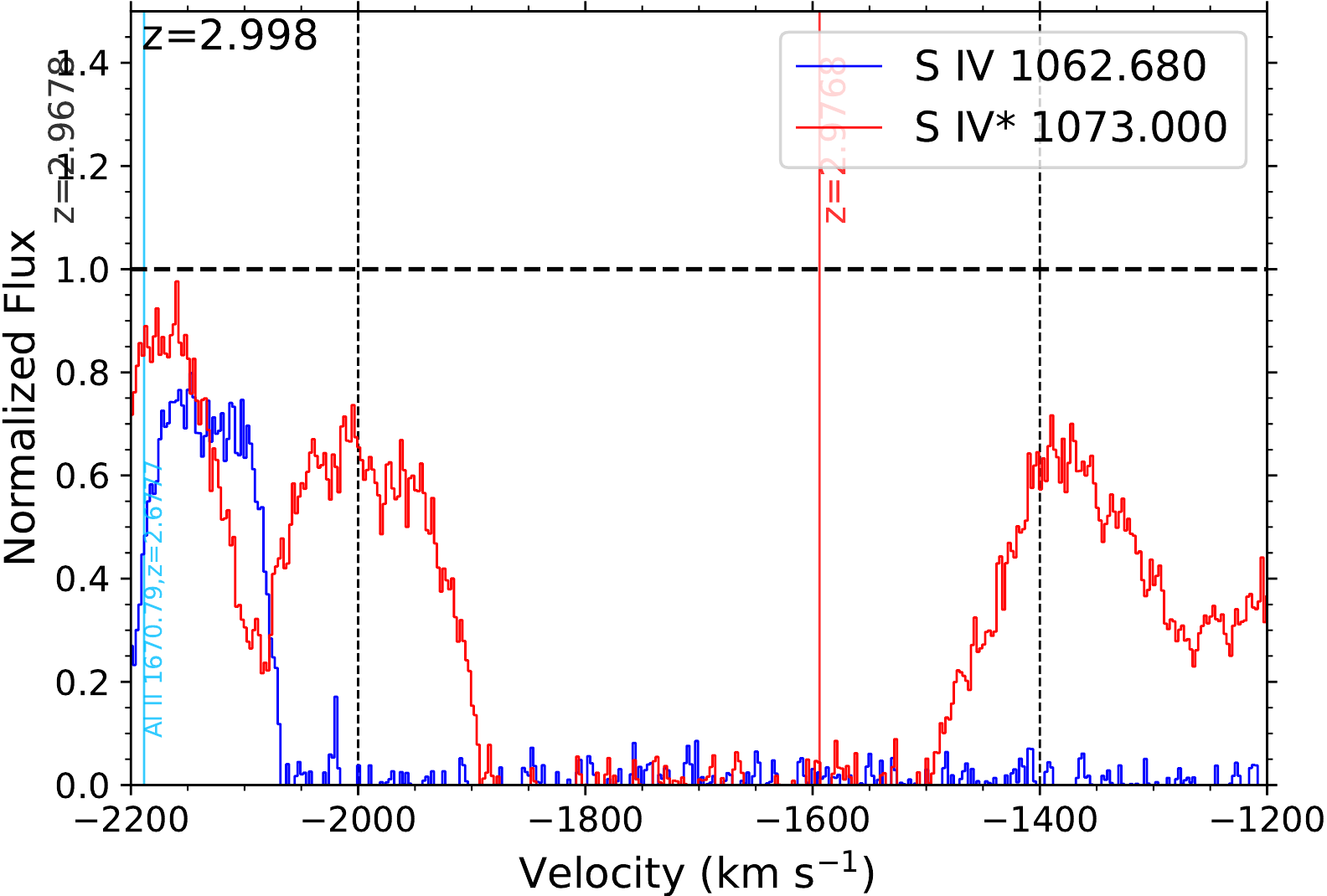}}\par
    \end{multicols}
    \begin{multicols}{3}
    \subcaptionbox{\ion{Si}{ii}\label{fig:SiII}}{\includegraphics[width=0.33\textwidth]{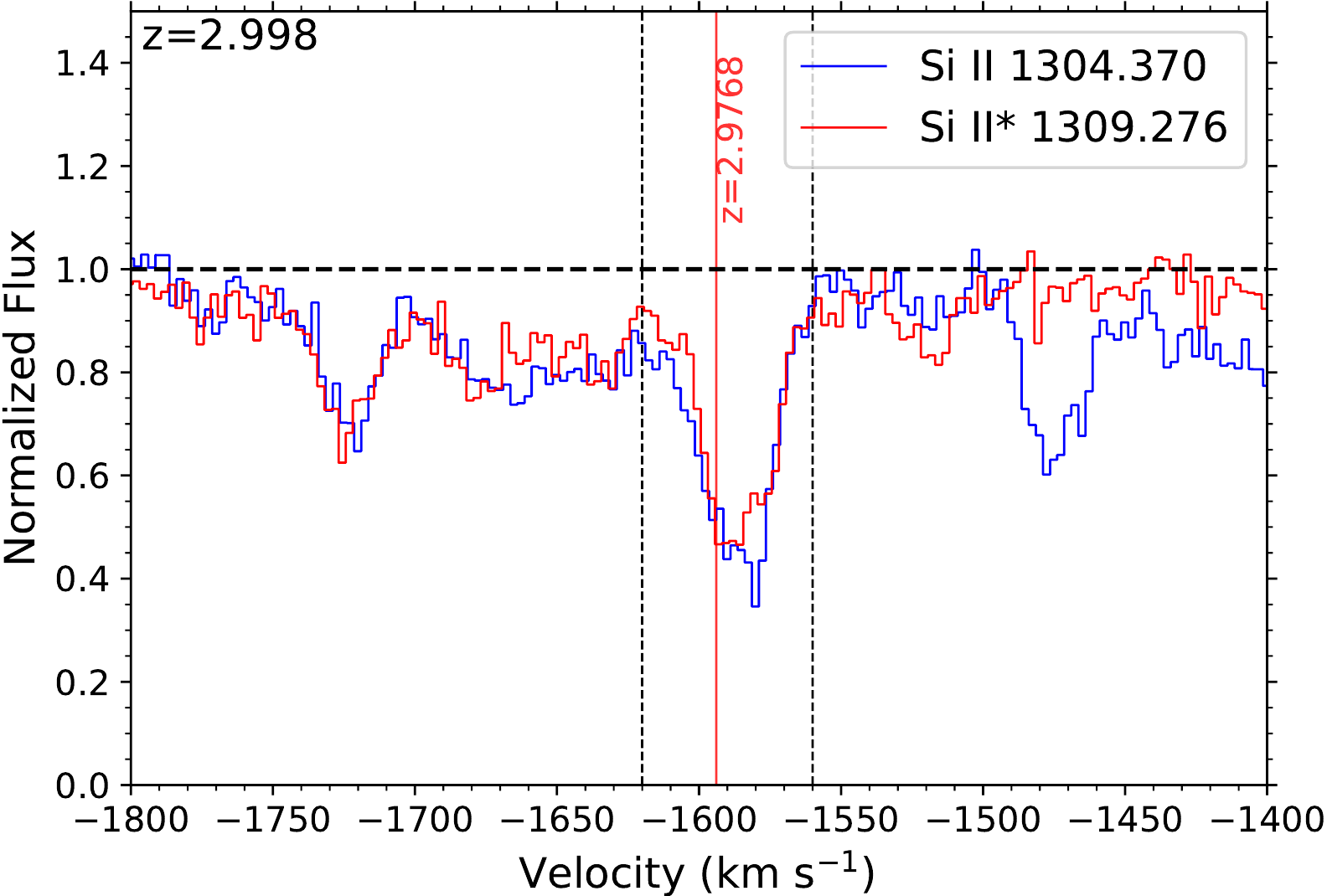}}\par
    \subcaptionbox{\ion{Fe}{ii}\label{fig:FeII}}{\includegraphics[width=0.33\textwidth]{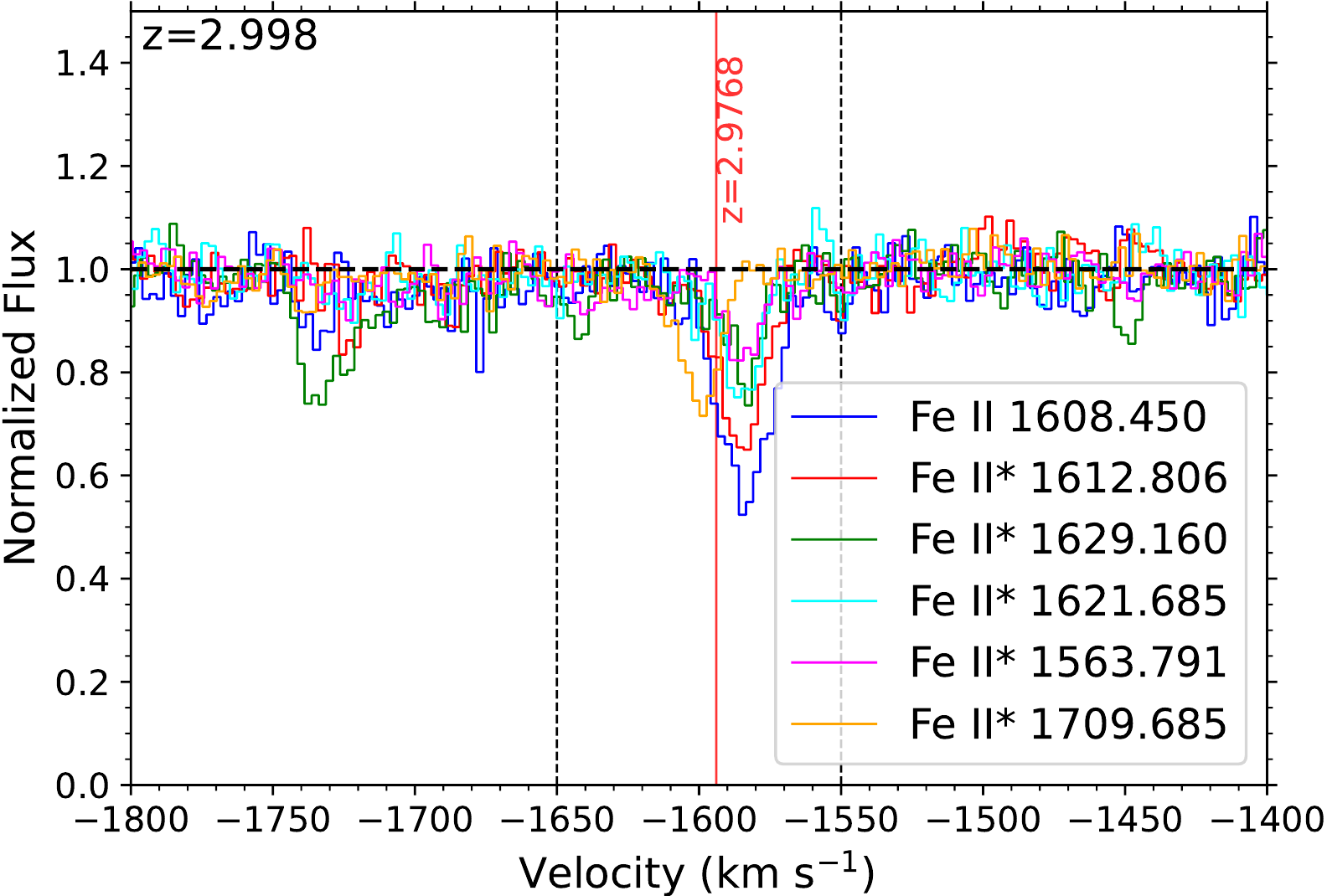}}\par
    \subcaptionbox{\ion{C}{ii}\label{fig:FeII}}{\includegraphics[width=0.33\textwidth]{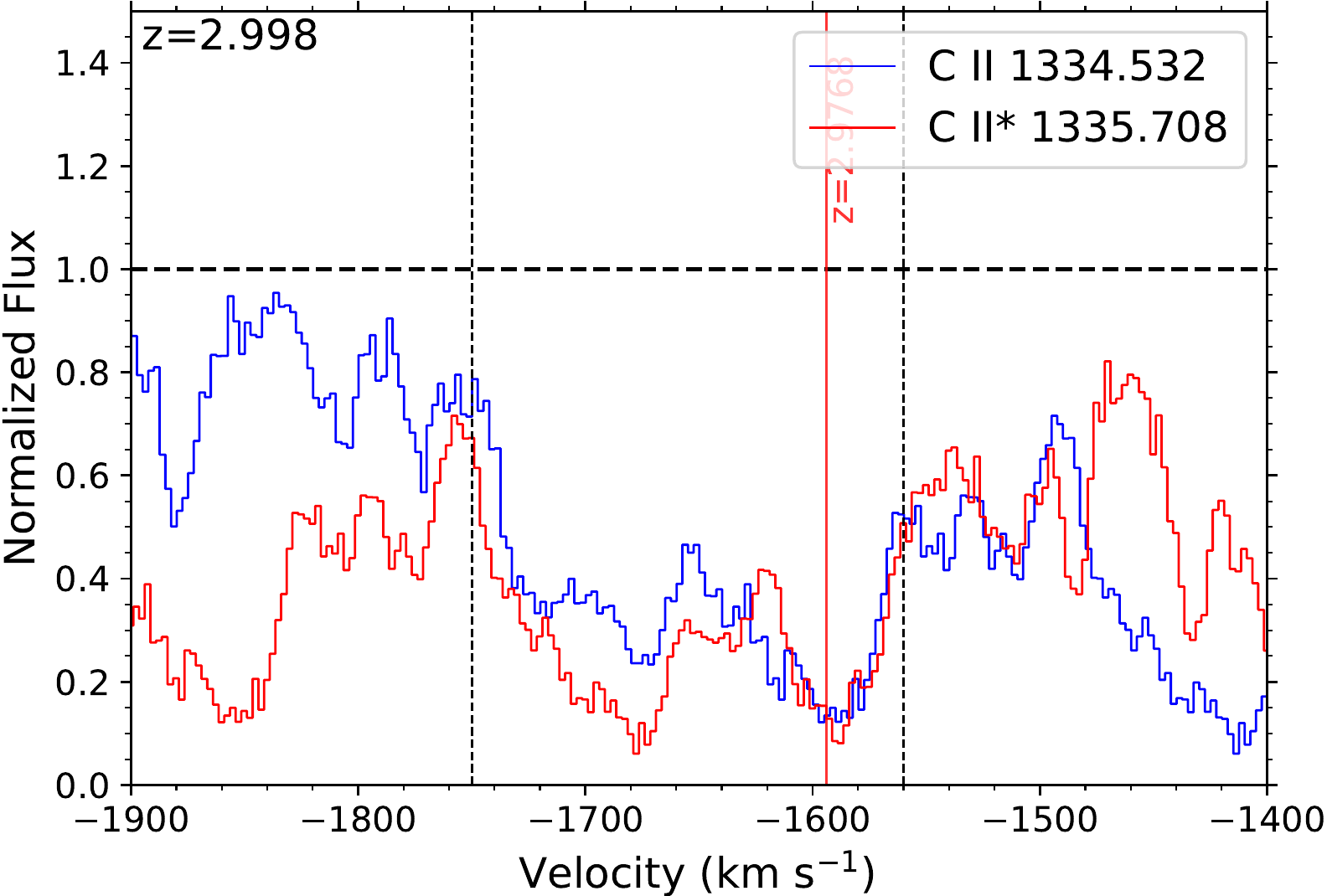}}\par
    \end{multicols}
\caption{Normalized spectra plotted in velocity space for ions in S1 that display excited states. The format of the plots are the same as in Fig.~\ref{fig:vcut}.}
\label{fig:vcut_excited}
\end{figure*}

The outflows show absorption of low ionization species such as \ion{Si}{ii}, \ion{Fe}{ii}, and \ion{O}{i}, and others including \ion{S}{iv}, \ion{C}{iv}, \ion{Si}{iv}, \ion{H}{i}, and \ion{Al}{iii}. The focus of this paper's analysis is on S1, as it displays troughs of excited states of \ion{Si}{ii}, \ion{Fe}{ii}, and \ion{O}{i}, which allowed us to find its $n_e$, and by extension, $R$. We converted the normalized spectrum from wavelength space to velocity space using the quasar's systemic redshift, as shown in the plots of Figs.~\ref{fig:vcut} and \ref{fig:vcut_excited}.\par
For our time variability analysis, we have retrieved SDSS spectra from MJD=51791, 52203, 55477, 56956, and 57688. We corrected the spectra for galactic reddening and extinction with $E(B-V)=0.0253$ \citep{2011ApJ...737..103S} and the extinction model by \citet{1999PASP..111...63F}. The SDSS spectra have resolutions of $R\approx2000$. More details on the SDSS spectra are in Table~\ref{table:sdss_spectra}.\par

\begin{table}
	\centering
	\caption{SDSS Spectra Information}
	\label{table:sdss_spectra}
	\begin{tabular}{lccccc} 
	\hline\hline
	MJD&Spectrograph&Plate&Fiber&Observed Date&Coverage (\AA)\\
	\hline
\text{51791}&SDSS&387&246&Sep. 4, 2000&3824--9215\\
\text{52203}&SDSS&685&317&Oct. 21, 2001&3820--9202\\
\text{55477}&BOSS&4216&410&Oct. 10, 2010&3565--10325\\
\text{56956}&BOSS&7850&555&Oct. 26, 2014&3604--10394\\
\text{57688}&BOSS&9208&969&Oct. 27, 2016&3608--10334\\
	\hline
	\end{tabular}
\end{table}

\section{Analysis} \label{sec:analysis}
\subsection{Ionic Column Density} \label{subsec:coldensity}
We found the ionic column densities ($N_{ion}$) of S1 as our first step in finding the physical characteristics of the outflow. We employed two different methods in measuring the column densities: assuming the apparent optical depth (AOD) of a uniform outflow covering a homogeneous source \citep{Savage1991}; and the partial covering (PC) method assuming a partially covered source \citep{1997ASPC..128...13B,1999ApJ...524..566A,Arav1999}.\par
While the AOD method is convenient in its simplicity, the PC method can provide us with more accurate measurements of ions with doublets of transition lines, taking into account effects such as non-black saturation \citep[e.g.,][]{2011ApJ...739....7E,2012ApJ...751..107B}. This is done by finding a velocity dependent covering factor for each ion \citep{DeKool2002,2005ApJ...620..665A}. A more detailed description of these methods can be found in \citet{2022Byun} Section 3.1, which used the same methods to analyze the UVES spectrum of the quasar J024221.87+004912.6. We also used an inhomogeneous power law (PL) covering factor for the \ion{P}{v} and \ion{Al}{iii} doublets to improve column density measurements, as done in previous work \citep[e.g.,][]{DeKool2002,2012ApJ...751..107B,2018ApJ...858...39X}. We incorporated the difference between the PL and PC values as the lower or upper errors for the adopted column density, depending on whether the PL result was smaller or larger than that from the PC method.\par
For each ion, we chose an integration range covering a visible absorption feature (see Figs.~\ref{fig:vcut}, \ref{fig:vcut_excited}). Some ions, such as \ion{N}{v} and \ion{C}{iv}, exhibit blending between the red and blue troughs of their doublets (see plots b and f of Fig.~\ref{fig:vcut}). To minimize the effects of blending, we selected an integration range that avoids an overlap of the red and blue features, and in this paper report a lower limit of the column density based on the AOD assumption. While we show both ground and excited state troughs of \ion{S}{iv} in Fig.~\ref{fig:vcut_excited}, since \ion{S}{iv} $\lambda1062$ is contaminated by the damped Lyman-$\alpha$ trough, we only measured the column density of \ion{S}{iv}* $\lambda1073$ and thus report a lower limit. While \ion{C}{ii} absorption is visibly present as seen in Fig.~\ref{fig:fluxplot}, the blending between the excited and resonance troughs, as well as contamination from intervening absorption, made it difficult to isolate the troughs. As such, while we computed a lower limit of the total column density of \ion{C}{ii}, we have excluded it from the electron number density calculation in Subsection \ref{subsec:edensity}.\par
The column density measurements are shown in Table~\ref{table:coldensity}. Note that the high column density of \ion{H}{i} comes from a measurement based on the Lyman 9 line, which has a wavelength of $\lambda=920.963\text{ \AA}$ (see plot a of Fig.~\ref{fig:vcut}). The errors in column density have been propagated from the error in normalized flux, binned in $\Delta v= 10 \text{km s}^{-1}$ wide bins. A conservative $20\%$ error has been added in quadrature to the adopted values used for photoionization analysis to take into account the uncertainty in the modeled continuum \citep{2018ApJ...858...39X}. Note that most of the adopted values are lower limits, as we have used the AOD method for most of the ions in the system. We also took AOD measurements of the visible absorption features of S2--S5. The lower limits of \ion{H}{i} column densities of S2--S4, measured from the Lyman-$\alpha$ lines, range between $\sim400-500 \times 10^{12}\text{ cm}^{-2}$, which is several orders of magnitude smaller than the measured \ion{H}{i} column density of S1. The \ion{C}{iv} of S5 has a lower limit of $N(\text{\ion{C}{iv}})>520\times10^{12}\text{ cm}^{-2}$.\par
\begin{table*}
	\centering
	\caption{J2357-0048 column densities from UVES observations}
	\label{table:coldensity}
	\renewcommand{\arraystretch}{1.4}
	\begin{tabular}{lccccccccc} 
	\hline\hline
	Troughs&Wavelength\text{(\AA)}&AOD&PC&PL&Adopted&$Z_\odot$-1&$Z_\odot$-2&$4.68Z_\odot$-1&$4.68Z_\odot$-2\\
	\hline
	\multicolumn{6}{c}{S1, $v=-1600\text{ km s}^{-1}$}\\
	\hline
    \text{\ion{H}{i}} &920.963&$141,700_{-7700}^{+18400}$  &&&$141,700^{+29,400}_{-33,800}$&16.4&3.50&8.0&2.92\\
    \text{\ion{N}{iii} total} &&$14,390_{-140}^{+830}$&&&$>14,390_{-2040}$&10.0&1.74&23.87&10.45\\
    \text{\ion{N}{iii} 0} &989.799&$7220_{-110}^{+700}$&&&\\
    \text{\ion{N}{iii}* 174} &991.577&$7270_{-90}^{+450}$&&&\\
    \text{\ion{N}{v}} &1238.820 \& 1242.800&$8640_{-30}^{+30}$&&&$>8640_{-1730}$&1.85&10.41&0.62&2.92\\
    \text{\ion{O}{i} total} &&$260_{-7}^{+7}$&&&$>260_{-30}$&0.03&0.75&0.20&0.85\\
    \text{\ion{O}{i} 0} &1302.171&$100_{-4}^{+4}$&&&\\
    \text{\ion{O}{i}* 158} &1304.861&$90_{-4}^{+4}$&&&\\
    \text{\ion{O}{i}* 227} &1306.032&$70_{-4}^{+4}$&&&\\
    \text{\ion{P}{v}} &1117.977 \& 1128.008&$677_{-6}^{+7}$&$740_{-10}^{+10}$&$510_{-10}^{+60}$&$510_{-100}^{+230}$&0.37&1.10&0.17&0.32\\
    \text{\ion{S}{iv}*} &1073.000&$14,330_{-180}^{+1710}$&&&$>14,330_{-2870}$&2.67&1.65&3.75&3.44\\
    \text{\ion{S}{vi}} &933.378 \& 944.523&$6870_{-120}^{+360}$&&&$>6870_{-1380}$&0.66&1.85&0.39&1.16\\
    \text{\ion{C}{ii} total} &&$1100_{-10}^{+10}$  &&&$>1100_{-220}$&5.27&1.48&11.66&3.18\\
    \text{\ion{C}{ii} 0} &1334.532&$490_{-5}^{+5}$  &&&&&&&\\
    \text{\ion{C}{ii}* 63} &1335.708&$610_{-6}^{+6}$  &&&&&&&\\
    \text{\ion{C}{iii}} &977.020&$790_{-10}^{+100}$  &&&$>790_{-160}$&815&173&498&293\\
    \text{\ion{C}{iv}} &1548.195 \& 1550.770&$8770_{-20}^{+530}$  &&&$>8770_{-1750}$&60.1&115&7.01&12.39\\
    \text{\ion{Si}{ii} total} && $230_{-4}^{+4}$&&&$>230_{-50}$&4.46&1.03&128&21.04\\
    \text{\ion{Si}{ii} 0} &1304.370& $120_{-3}^{+3}$&&&\\
    \text{\ion{Si}{ii}* 287} &1309.276& $110_{-2}^{+3}$&&&\\
    \text{\ion{Si}{iv}} &1393.755 \& 1402.770& $3320_{-20}^{+260}$&&&$>3320_{-660}$&18.0&3.75&43.2&41.29\\
    \text{\ion{Al}{ii}} &1670.787& $10_{-0.2}^{+0.2}$&&&$>10_{-2}$&8.3&2.0&51.2&6.16\\
    \text{\ion{Al}{iii}} &1854.716 \& 1862.790& $163_{-1.4}^{+1.4}$&$218_{-3}^{+57}$&$391_{-3}^{+76}$&$390_{-170}^{+110}$&1.97&0.22&5.46&2.37\\
    \text{\ion{Fe}{ii} total} &&$260_{-10}^{+10}$&&&$>260_{-50}$&0.01&0.44&0.40&1.27\\
    \text{\ion{Fe}{ii} 0} &1608.450& $60_{-3}^{+3}$&&&\\
    \text{\ion{Fe}{ii}* 385} &1621.685& $40_{-4}^{+5}$&&&\\
    \text{\ion{Fe}{ii}* 668} &1629.160& $36_{-4}^{+5}$&&&\\
    \text{\ion{Fe}{ii}* 1873} &1612.806& $70_{-5}^{+6}$&&&\\
    \text{\ion{Fe}{ii}* 2430} &1563.791& $17_{-2}^{+2}$&&&\\
    \text{\ion{Fe}{ii}* 13474} &1709.685& $35_{-2}^{+3}$&&&\\
		\hline
	\end{tabular}
	\tablewidth{\linewidth}
	\tablecomments{Column densities have been calculated via numerical integration over $\Delta v = 10 \text{km s}^{-1}$ bins. Energy states of ions are noted with energies in units of cm$^{-1}$. The wavelengths listed for each ion are the transition rest wavelengths in \AA. Most of the adopted values are lower limits due to being AOD measurements. The four rightmost columns show the ratios between modeled and adopted column densities, for the solar one-phase, solar two-phase, supersolar one-phase, and supersolar two-phase solutions, respectively. Units of column density are in $10^{12}\text{cm}^{-2}$.}
\end{table*}

\subsection{Photoionization Analysis} \label{subsec:photoionization}
\begin{figure}
    \centering
    \subcaptionbox{One-phase solution\label{fig:onephase}}{\includegraphics[width=\linewidth]{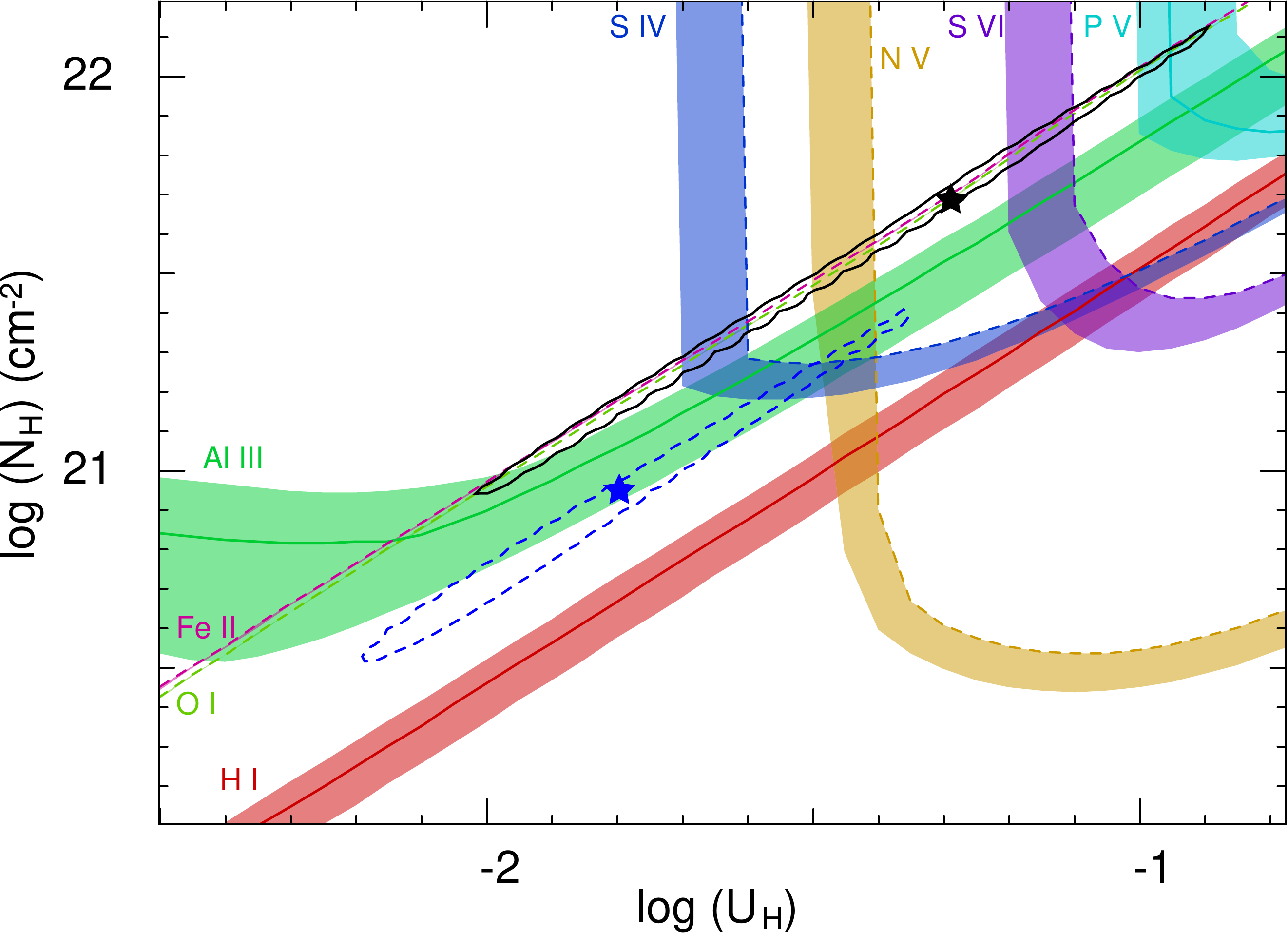}}\par
    \subcaptionbox{Two-phase solution\label{fig:twophase}}{\includegraphics[width=\linewidth]{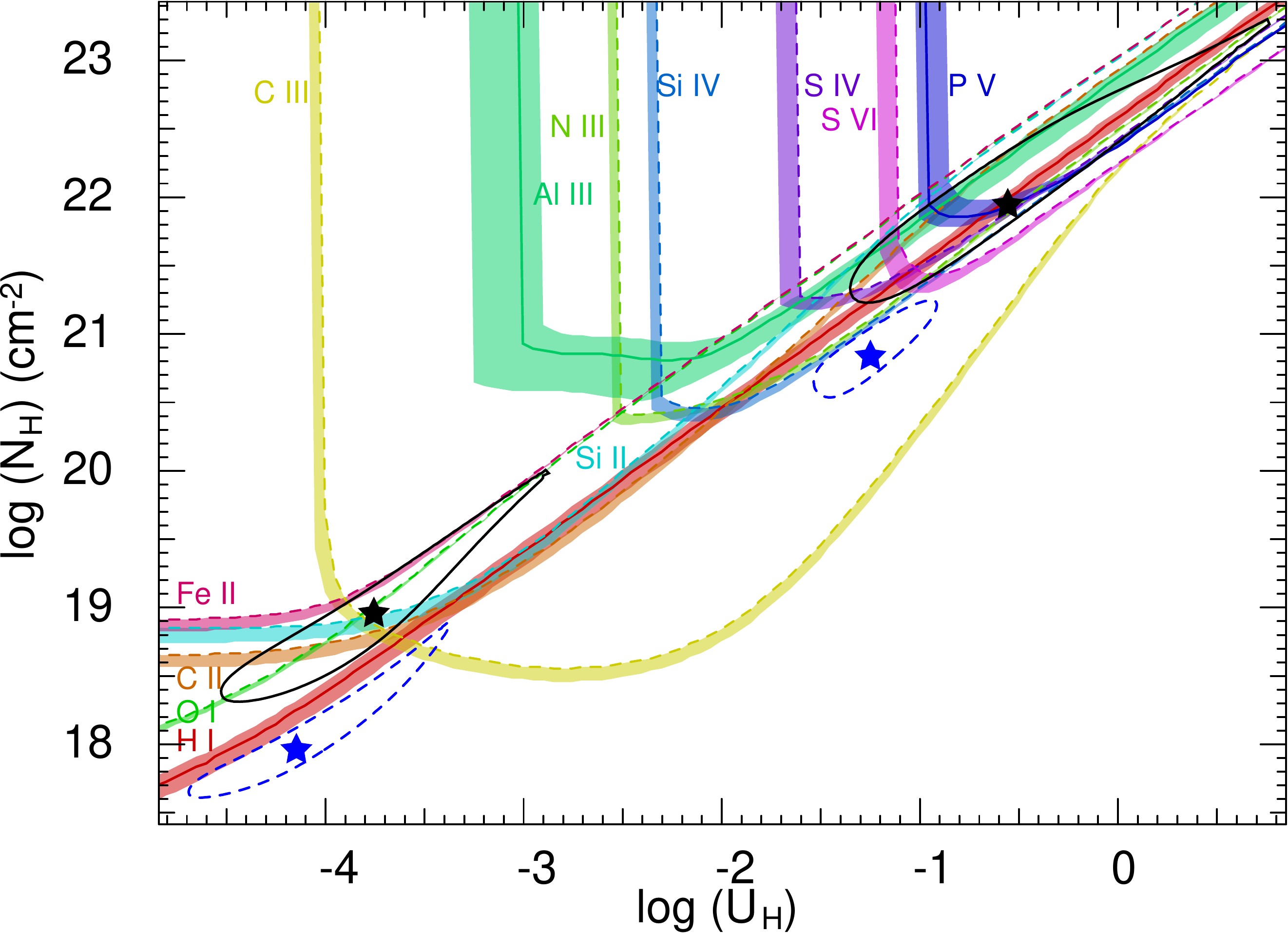}}\par
    \caption{Plot of $\log{N_H}$ vs. $\log{U_H}$ for S1. The colored lines represent the $N_H$ and $U_H$ values allowed by the measured column densities of ions. Solid and dashed lines represent measurements and lower limits, respectively. The colored bands attached to the lines show the uncertainties in ionic column densities. The black stars and ellipses represent the $N_H$ and $U_H$ solutions, and the $1\sigma$ range, assuming solar metallicity. The blue stars and blue dashed ellipses represent the solutions and $1\sigma$ range assuming supersolar metallicity \citep[$Z=4.68Z_\odot$,][]{2008A&A...478..335B,2020ApJS..247...41M}. Note that the lower limits of ions such as \ion{C}{iv} and \ion{N}{v} are satisfied by any point above their respective colored bands. The high-ionization phase of the two-phase solution in plot b satisfies the lower limits from \ion{S}{iv}, \ion{S}{vi}, \ion{C}{iv}, and \ion{N}{v}, making the two-phase solution a better fit to the constraints than the one-phase solution shown in plot a. The HE0238 SED is assumed.}
    \label{fig:nvuplot}
\end{figure}

We followed the method of previous works to find the ionization parameter ($U_H$) and the Hydrogen number density ($N_H$) of the outflow \citep{2019ApJ...876..105X,2020ApJS..247...39M,2020ApJS..249...15M,2022Byun}. Using Cloudy \citep[][version c17.00]{2017RMxAA..53..385F}, we created a grid of photoionization models in order to find the values of $N_H$ and $U_H$ that best fit the ionic column density measurements. Assuming the spectral energy distribution (SED) of quasar HE0238-1904 (hereafter HE0238) \citep{2013MNRAS.436.3286A}, we created a grid of models with varying values of $N_H$ and $U_H$. The $N_H$ and $U_H$ values used for each model determines the column density of each ion, which we compared with the measured values in Table \ref{table:coldensity}. A $\chi^2$ analysis has given us the closest matching model to the measurements, as shown in Fig.~\ref{fig:nvuplot}.\par
We assumed solar metallicity and searched for a one-phase solution (see plot a of Fig.~\ref{fig:nvuplot}), and here report the best fitting $N_H$ and $U_H$ in Table \ref{table:energetics}. The solution displays a poor fit to the constraints given by ions such as \ion{P}{v}, \ion{S}{vi}, and \ion{N}{v}. To improve the models' fit with the data, we have calculated one-phase (plot a of Fig.~\ref{fig:nvuplot}) and two-phase (plot b of Fig.~\ref{fig:nvuplot}) solutions with both solar and supersolar \citep[$Z=4.68Z_\odot$,][]{2008A&A...478..335B,2020ApJS..247...41M} metallicities, and computed the physical parameters as shown in Table \ref{table:energetics}. The ratios between modeled and measured column densities can be found in Table~\ref{table:coldensity}. \par
 For the two-phase solutions, we attributed the column densities of \ion{Fe}{ii} and \ion{O}{i} to the low-ionization phase, as it was more in line with the constraints given by the measured column densities. The $\log{n_e}$ values (as estimated in Section \ref{subsec:edensity}) for the high-ionization phase were calculated assuming that the low and high-ionization phases shared the same distance from the central source. Note that the two-phase solutions, both for solar and supersolar metallicity, yield better fits to the column density measurements (see Fig.~\ref{fig:nvuplot}).\par
\subsection{Electron Number Density} \label{subsec:edensity}
The electron number density ($n_e$) of the outflow can be found through comparing the column densities of different energy states of ions \citep[e.g.,][]{2009ApJ...706..525M}. S1 has absorption troughs of the resonance and excited states of \ion{Si}{ii}, \ion{O}{i}, and \ion{Fe}{ii}, which we could potentially use for our $n_e$ measurement. We used the CHIANTI 9.0.1 Database \citep{1997A&AS..125..149D,Dere_2019} for this task, as it models the ratio between energy states dependent on $n_e$ based on collisional excitation.\par
The \ion{Si}{ii} ground and excited states display troughs that are nearly identical in depth (See plot d in Fig.~\ref{fig:vcut_excited}), suggesting non-black saturation. As it is difficult to find a concrete constraint on the ratio between the two states, we excluded \ion{Si}{ii} from this process. Finding the ratios between excited and ground states of \ion{Fe}{ii} and \ion{O}{i} gave unphysical values, suggesting saturation of the ground states of \ion{Fe}{ii} and \ion{O}{i} as well. Thus, we found the ratios between excited states instead, as shown in Fig.~\ref{fig:ratioplot}. We then found the weighted mean of the $\log{n_e}$ values from the different ratios, following the linear model method described by \citet{2003sppp.conf..250B}. The resulting value of $\log{n_e}$ can be seen in Table \ref{table:energetics}.\par
While we excluded \ion{Si}{ii} from the process of finding $\log{n_e}$, we confirmed that the lower limit of the \ion{Si}{ii} column density was satisfied by the combination of the $N_H$, $U_H$, and $n_e$ parameters found by using Cloudy models. The solar one-phase solution estimates $\log{N(\text{\ion{Si}{ii}})}\approx15.01$, while the low ionization phase of the solar two-phase solution predicts $\log{N(\text{\ion{Si}{ii}})}\approx14.37$, both of which are above the lower limit reported in Table \ref{table:coldensity}. While a comparison with an analysis of \ion{C}{iii}* would have been helpful for additional confirmation of $n_e$ \citep[e.g.][]{2005ApJ...631..741G,2012ApJ...758...69B,2015A&A...577A..37A,2018ApJ...866....7L}, we were unable to identify the absorption troughs of the \ion{C}{iii}* 1175 \AA\text{ } multiplet, due to their indistinguishability with the Lyman-alpha forest lines (See Figure~\ref{fig:fluxplot}). We modeled the column densities of the different energy levels with Cloudy using the $N_H$, $U_H$, and $n_e$ values corresponding to each solution (See Table~\ref{table:c3}). The ratio between the J=2 energy state and the J=0 state ranges between 0.003 and 1.86.

\section{Results} \label{sec:results}
\subsection{Distance and Kinetic Luminosity of the Outflow}\label{subsec:energetics}
The electron number density allows us to find the distance of the outflow from its central source, based on the following definition of the ionization parameter:
\begin{equation}
    \label{eq:UH}
    U_H \equiv \frac{Q_H}{4\pi R^2 n_H c}
\end{equation}
where $Q_H$ is the rate of Hydrogen ionizing photons, R is the distance of the outflow from the quasar, and $n_H$ is the Hydrogen number density. Since $n_e \approx 1.2 n_H$ for highly ionized plasma \citep{2006agna.book.....O}, and we found the values of both $U_H$ and $n_e$ in Section \ref{sec:analysis} (See Table \ref{table:energetics}), we could solve the equation for R once the value of $Q_H$ was determined.\par
Following the method of previous works \citep[e.g.][]{2020ApJS..247...39M,2022Byun}, we scaled the HE0238 SED to match the continuum flux at rest wavelength $\lambda=1350$ \text{\AA \ } from the SDSS observation on October 21, 2001 ($F_\lambda= 7.27^{+0.82}_{-0.82}\times10^{-17}\text{ erg s}^{-1}\text{ cm}^{-2} \AA^{-1}$), and integrated over the scaled SED for energies above 1 Ryd, finding bolometric luminosity $L_{Bol}=1.8^{+0.2}_{-0.2}\times10^{47}\text{ erg s}^{-1}$ and $Q_H=1.13^{+0.13}_{-0.13}\times10^{57}\text{ s}^{-1}$. The outflow distance calculated from these values is shown in Table \ref{table:energetics}.\par
Assuming an outflow with the geometry of an incomplete spherical shell, we calculated the mass flow rate and kinetic luminosity with the following equations \citep{2012ApJ...751..107B}:
\begin{equation}
    \label{eq:Mdot}
    \dot{M} \simeq 4\pi\Omega R N_H \mu m_p v
\end{equation}
\begin{equation}
    \label{eq:Ekdot}
    \dot{E}_k \simeq \frac{1}{2} \dot{M} v^2
\end{equation}
where $\Omega$ is the global covering factor (fraction of the solid angle covered by the outflow), $\mu = 1.4$ is the mean atomic mass per proton, $m_p$ is the proton mass, and $v$ is the outflow velocity. We assumed $\Omega = 0.2$, which is the portion of quasars from which \ion{C}{iv} BALs are detected \citep{2003AJ....125.1784H}. Calculating the kinetic luminosity yielded values ranging from $\log{\dot{E}_K}=43.06^{+0.21}_{-0.36}\text{ [erg s}^{-1}]$ for the one-phase solution assuming solar metallicity, to $\log{\dot{E}_K}=45.55^{+0.90}_{-0.32}\text{ [erg s}^{-1}]$ for the two-phase solution assuming solar metallicity. Results based on the other solutions can be found in Table \ref{table:energetics}.

\subsection{Changes in the High Velocity Mini-BAL Trough (S5)}\label{subsec:velocityshift}
\begin{figure}
    \centering
    \subcaptionbox{\label{fig:BAL}}{\includegraphics[width=\linewidth]{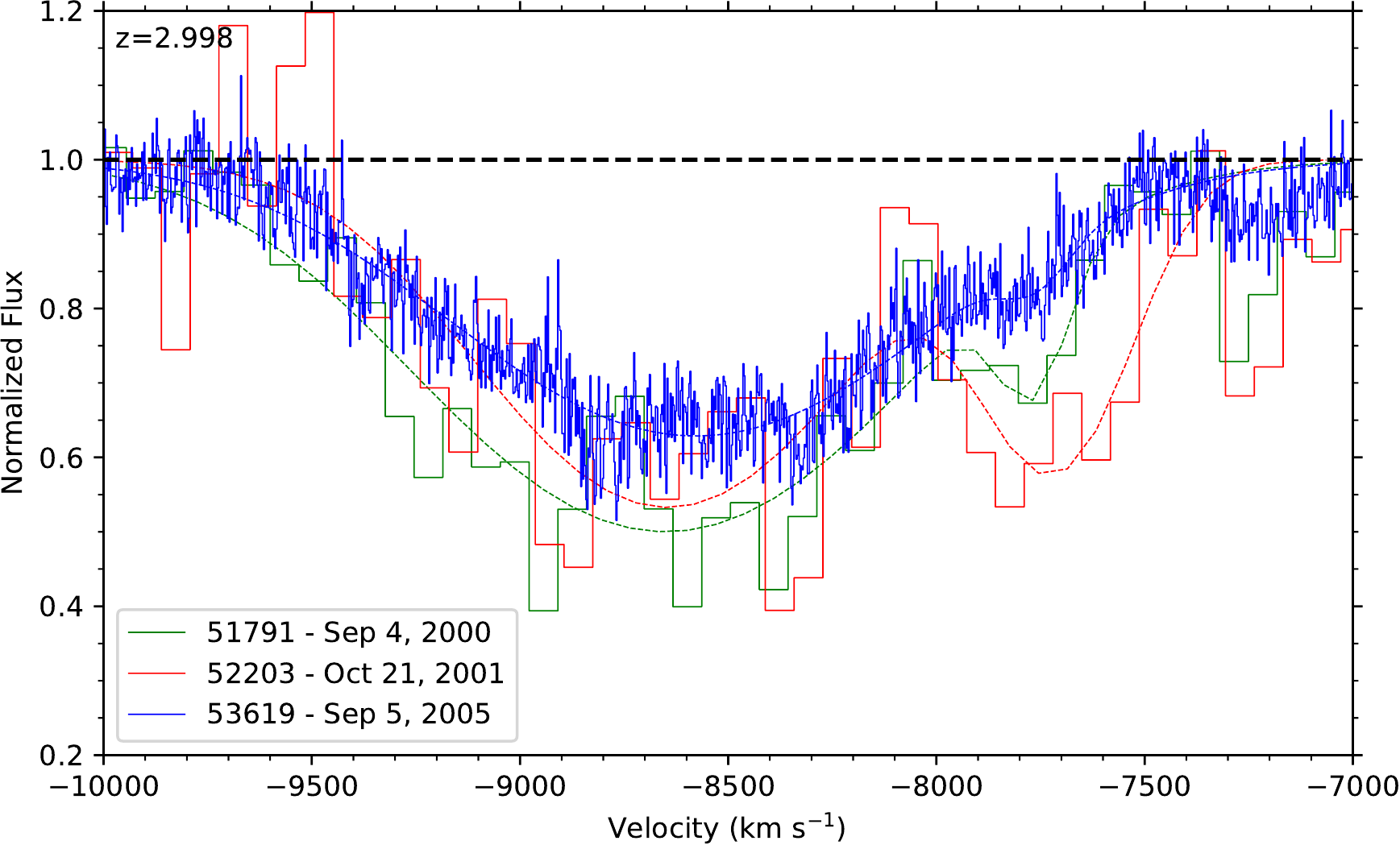}}\par
    \subcaptionbox{\label{fig:BAL_gaussian}}{\includegraphics[width=\linewidth]{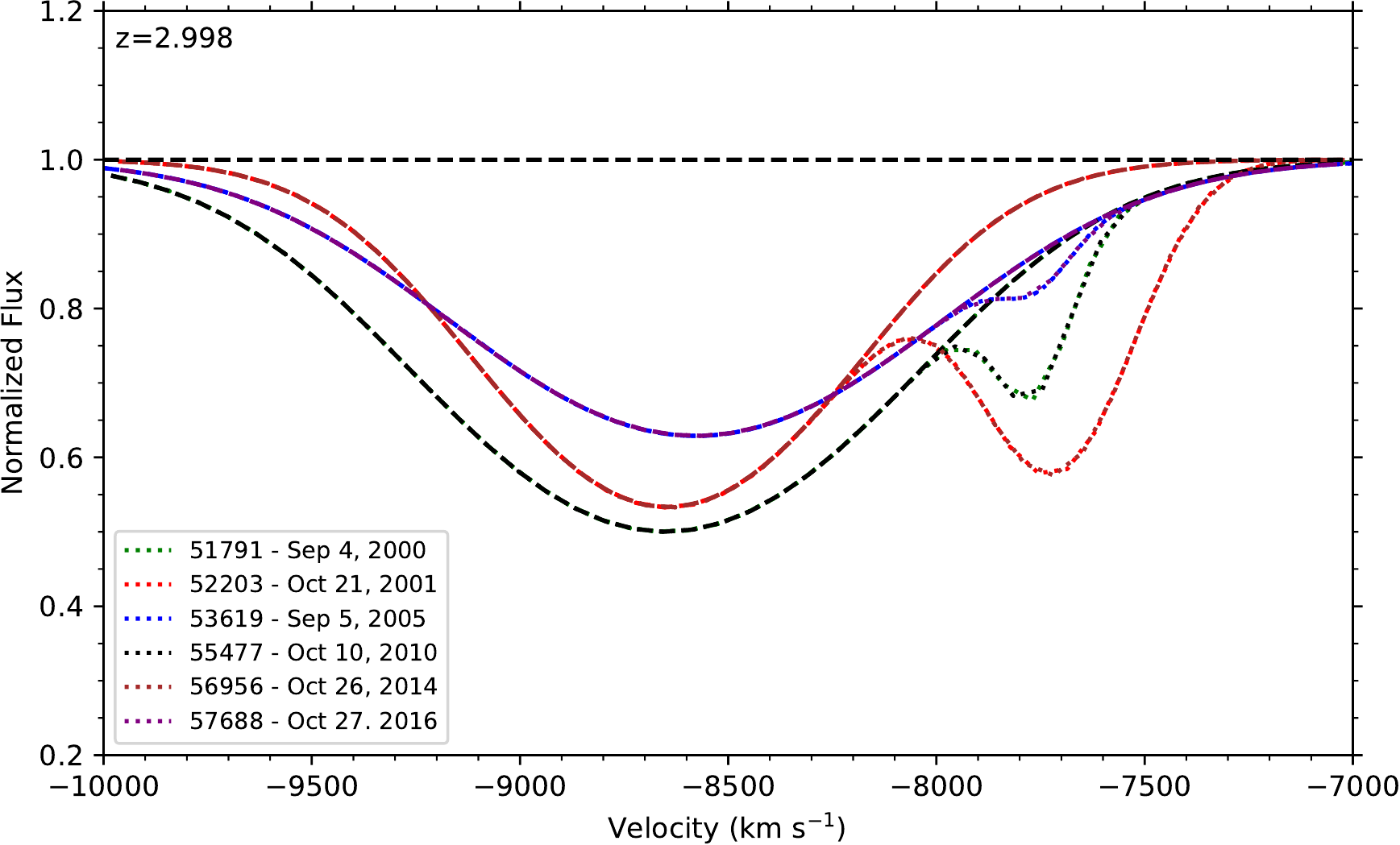}}\par
    \caption{Normalized flux vs. velocity of the S5 \ion{C}{iv} mini-BAL at three observational epochs. A best fit of two Gaussians was used to model the absorption. (a) shows the models over the data of three epochs, and (b) shows the Gaussians of all six epochs.}
    \label{fig:bigBALvelocities}
\end{figure}

We examined the time-variability of the \ion{C}{iv} mini-BAL of S5. We adopted the method of \citet{2022Byun}, fitting two Gaussian profiles, one broad and one narrow, with the absorption. We added a constraint to the centroid velocity of the narrow Gaussian ($-7900\text{ km s}^{-1}<v_c<-7600\text{ km s}^{-1}$), as there was a distinct absorption feature found at $v\approx-7800\text{km s}^{-1}$. As shown in Fig.~\ref{fig:bigBALvelocities}, the smaller absorption feature grows between the observations in 2000 and 2001, but nearly vanishes by 2005. The larger feature becomes gradually shallower. More information of the Gaussian fits can be found in Table \ref{table:epoch_velocity}.\par
\begin{figure}
    \centering
    \includegraphics[width=\linewidth]{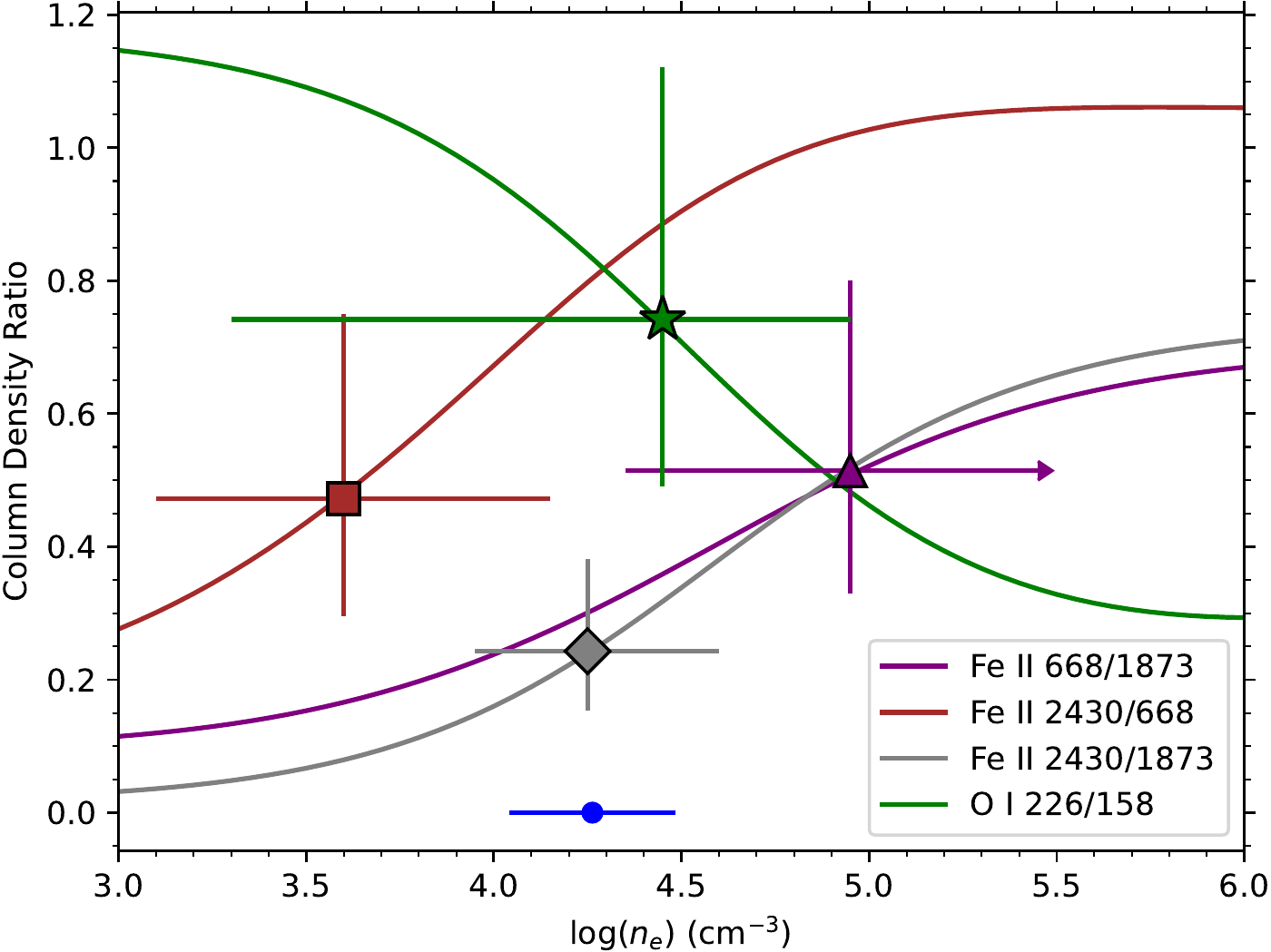}
    \caption{The ratio between the abundances of different energy states of \ion{O}{i} and \ion{Fe}{ii} in S1, assuming a temperature of 10,000 K. The curves represent the theoretical ratios modeled with CHIANTI. The crosses on the curves represent the upper and lower limits of the ratios based on the measured column densities, along with the associated $\log{n_e}$ values.The ratios are color coded and marked with shapes:  \ion{Fe}{ii}* 668/1873 with a triangle, \ion{Fe}{ii}* 2430/668 with a square, \ion{Fe}{ii}* 2430/1873 with a diamond, and \ion{O}{i}* 226/158 with a star. The rightward arrow on the purple cross of the \ion{Fe}{ii}* 668/1873 ratio shows that the data point is a lower limit, as the upper limit of the ratio exceeds the highest point of the curve. The blue dot and error bars show the weighted average of $\log{n_e}$ based on the four measured ratios, which we adopted for our analysis.}
    \label{fig:ratioplot}
\end{figure}
\begin{table*}
	\centering
	\caption{Physical Properties of J2357-0048 Outflow. Note that the high-ionization and low-ionization phases of the two-phase solutions are assumed to be co-spatial (at the same distance from the central source). The kinetic luminosity to bolometric luminosity ratios are included for comparison.}
	\label{table:energetics}
	\begin{tabular}{lcccccc}
	\hline\hline
	Solution &\text{Solar, one-phase}&\multicolumn{2}{c}{Solar, two-phase}&\text{Supersolar, one-phase}&\multicolumn{2}{c}{Supersolar, two-phase}\\
Phase&One-phase&Low&High&One-phase&Low&High\\
	\hline
\vspace{-0.2cm}$log(N_{\text{H}})$&\\\vspace{-0.2cm}
&$21.69^{+0.44}_{-0.75}$&$18.96^{+1.05}_{-0.64}$&$21.95^{+1.36}_{-0.72}$&$20.95^{+0.46}_{-0.43}$&$17.96^{+0.93}_{-0.35}$&$20.84^{+0.41}_{-0.30}$\\
$[\text{cm}^{-2}]$&\\
\hline
\vspace{-0.2cm}$log(U_{\text{H}})$&\\\vspace{-0.2cm} &$-1.29^{+0.39}_{-0.73}$&$-3.76^{+0.88}_{-0.77}$&$-0.56^{+1.32}_{-0.80}$&$-1.80^{+0.44}_{-0.39}$&$-4.15^{+0.77}_{-0.54}$&$-1.25^{+0.34}_{-0.29}$\\
$[\text{dex}]$&\\
\hline
\vspace{-0.2cm}$log(n_{\text{e}})$&\\\vspace{-0.2cm} &$4.3^{+0.2}_{-0.2}$&$4.3^{+0.2}_{-0.2}$&$1.1^{+1.2}_{-1.5}$&$4.3^{+0.2}_{-0.2}$&$4.3^{+0.2}_{-0.2}$&$1.4^{+0.9}_{-0.7}$\\
$[\text{cm}^{-3}]$&\\
\hline
\vspace{-0.2cm}$\text{Distance}$&\\\vspace{-0.2cm}&$640^{+890}_{-260}$&\multicolumn{2}{c}{$10900^{+16600}_{-7100}$}&$1140^{+780}_{-500}$&\multicolumn{2}{c}{$17100^{+16600}_{-10300}$}\\
$[\text{pc}]$&\\
\hline
\vspace{-0.2cm}$\dot M$&\\\vspace{-0.2cm}&$14^{+9}_{-8}$&\multicolumn{2}{c}{$4400^{+30700}_{-2270}$}&$47^{+29}_{-18}$&\multicolumn{2}{c}{$540^{+1190}_{-370}$}\\
$[M_{\odot} \text{yr}^{-1}]$&\\
\hline
\vspace{-0.2cm}$\dot M v$&\\\vspace{-0.2cm}&$14^{+9}_{-8}$&\multicolumn{2}{c}{$4400^{+30800}_{-2280}$}&$47^{+29}_{-18}$&\multicolumn{2}{c}{$540^{+1190}_{-370}$}\\
$[10^{34} \text{ ergs cm}^{-1}]$&\\
\hline
\vspace{-0.2cm}$log({\dot E}_K)$&\\\vspace{-0.2cm}&$43.06^{+0.21}_{-0.36}$&\multicolumn{2}{c}{$45.55^{+0.90}_{-0.32}$}&$43.57^{+0.21}_{-0.21}$&\multicolumn{2}{c}{$44.63^{+0.51}_{-0.50}$}\\
$[\text{erg s}^{-1}]$&\\
\hline
\vspace{-0.2cm}${\dot E}_K/L_{Edd}$&\\\vspace{-0.2cm}&$0.008^{+0.006}_{-0.004}$&\multicolumn{2}{c}{$2.35^{+16.80}_{-1.27}$}&$0.025^{+0.019}_{-0.011}$&\multicolumn{2}{c}{$0.29^{+0.67}_{-0.20}$}\\
$[\text{\%}]$&\\
\hline
\vspace{-0.2cm}${\dot E}_K/L_{Bol}$&\\\vspace{-0.2cm}&$0.006^{+0.004}_{-0.004}$&\multicolumn{2}{c}{$1.96^{+13.68}_{-1.02}$}&$0.021^{+0.013}_{-0.008}$&\multicolumn{2}{c}{$0.24^{+0.53}_{-0.16}$}\\
$[\text{\%}]$&\\
	\hline
	\end{tabular}
\end{table*}
\begin{table*}
	\centering
	\caption{Modeled \ion{C}{iii}* column densities of the J2357-0048 outflow, in units of cm$^{-2}$, along with the ratio between the energy states of J=2 and J=0.}
	\begin{tabular}{lcccc}
	    \hline\hline
	    Energy level&Solar,one-phase&Solar,two-phase&Supersolar,one-phase&Supersolar,two-phase\\
	    \hline
	    J=0&$3.66\times10^{13}$&$9.92\times10^{12}$&$1.25\times10^{11}$&$4.48\times10^9$\\
	    J=1&$2.96\times10^9$&$1.21\times10^6$&$1.26\times10^7$&$1.81\times10^3$\\
	    J=2&$5.98\times10^{13}$&$3.14\times10^{10}$&$2.33\times10^{11}$&$2.74\times10^{7}$\\
	    J=2/J=0&1.63&0.003&1.86&0.006\\
	    \hline
	\end{tabular}
	\label{table:c3}
\end{table*}
\begin{table*}
    \caption{Gaussian Fits of \ion{C}{iv} mini-BAL at Each Epoch}
    \begin{tabular}{lcccccccc}
        \hline\hline
        MJD&Date&$\Delta t_{Rest}$&$v_{n}$&$\Delta v_{n}$&$EW_n$&$v_{w}$&$\Delta v_{w}$&$EW_w$\\
        &&(days)&($\text{km s}^{-1}$)&($\text{km s}^{-1}$)&($\text{km s}^{-1}$)&($\text{km s}^{-1}$)&($\text{km s}^{-1}$)&($\text{km s}^{-1}$)\\
        \hline
        \text{51791}&Sep. 4, 2000&$0$&$-7770\pm{40}$&$0$&$45$&$-8650\pm{40}$&$0$&$690$\\
        \text{52203}&Oct. 21, 2001&193.9&$-7710\pm{50}$&$+60\pm{60}$&$180$&$-8650\pm{70}$&$0\pm{80}$&$500$\\
        \text{53619}&Sep. 5, 2005&860.6&$-7760\pm{14}$&$-10\pm{40}$&$12$&$-8577\pm{6}$&$+70\pm{40}$&$515$\\
        \text{55477}&Oct. 10, 2010&1735.4&$-7830\pm{60}$&$-60\pm{70}$&$15$&$-8650\pm{50}$&$+0\pm{70}$&$500$\\
        \text{56956}&Oct. 26, 2014&2431.7&$-7900\pm{5180}$&$-130\pm{5180}$&$6$&$-8600\pm{50}$&$+50\pm{70}$&$470$\\
        \text{57688}&Oct. 27, 2016&2776.3&$-7900\pm{200}$&$-10\pm{200}$&$15$&$-8590\pm{80}$&$+60\pm{90}$&$500$\\
        \hline
    \end{tabular}
    \label{table:epoch_velocity}
    \tablecomments{Table of the centroid velocity and equivalent width of the Gaussian profiles fit to the \ion{C}{iv} absorption of S5. $v_n$ and $v_w$ are the centroid velocities of the narrow and wide Gaussians. $EW_n$ and $EW_w$ are the equivalent widths of the Gaussians in velocity space. $\Delta v_n$, $\Delta v_w$, and $\Delta t_{Rest}$ are the changes in velocity and time in the quasar's rest frame from the MJD=51791 epoch.}
\end{table*}

\section{Discussion} \label{sec:discussion}
\subsection{AGN Feedback Contribution of the Outflow}
The kinetic luminosity ($\dot{E}_K$) of the outflow must be at least $\sim0.5\%$ \citep{2010MNRAS.401....7H} or $\sim5\%$ \citep{2004ApJ...608...62S} of the quasar's Eddington luminosity ($L_{Edd}$) to be a significant contributor to AGN feedback. To find the ratio, we first found the mass and the Eddington luminosity of the quasar. Following the method demonstrated by \citet{2006ApJ...641..689V}, we computed the mass of the black hole based on the FWHM of the \ion{C}{iv} emission in the SDSS spectrum. To account for the blueshift of \ion{C}{iv}, we adopted the correction method by \citet{2017MNRAS.465.2120C}. This yields a mass of $M_{BH}=1.16^{+0.40}_{-0.30}\times10^9M_\odot$, corresponding to an Eddington luminosity of $L_{Edd}=1.5^{+0.5}_{-0.4}\times10^{47}\text{ erg s}^{-1}$. Our one-phase solutions yield $\dot{E}_K/L_{Edd}=0.008^{+0.006}_{-0.004}\%$ for solar metallicity and $\dot{E}_K/L_{Edd}=0.025^{+0.019}_{-0.011}\%$ for supersolar metallicity, while the two-phase solutions yield $\dot{E}_K/L_{Edd}=2.35^{+16.80}_{-1.27}\%$ for solar metallicity and $\dot{E}_K/L_{Edd}=0.29^{+0.67}_{-0.20}\%$ for supersolar metallicity (See Table \ref{table:energetics}). All but the two-phase solar metallicity solution are insufficient to contribute to AGN feedback. The $2.35\%$ ratio from the two-phase solar metallicity solution is above the $\sim0.5\%$ cutoff by \citet{2010MNRAS.401....7H}. This is in contrast with the outflow of SDSS J024221.87+004912.6, another SQUAD quasar analyzed by \citet{2022Byun}, which exceeded 5\% of the quasar's Eddington luminosity. Other quasar absorption outflows in the literature have $\dot{E}_k/L_{Edd}$ values ranging from $0.001\%$ \citep{2015MNRAS.454..675C} to an upwards of $\sim10\%$ \citep{2015MNRAS.450.1085C}. A larger sample size of analyzed objects may be needed for a statistical analysis. \par

\subsection{The Time Variability of S5}
The most notable change in S5 over time has been of the width and depth of the narrow feature at $v=-7800\text{ km s}^{-1}$, which disappeared in the spectrum of September 2005. The variability of the broader feature afterwards has been minimal, with the equivalent width staying at $\sim500\text{ km s}^{-1}$.\par
One possible explanation for the variability is that there has been a change in photoionization over the different epochs, affecting the depth of the narrow subcomponent \citep[e.g.][]{,2020ApJS..247...40X,2022Byun}. While other works involve the acceleration of outflow systems due to significant velocity shifts \citep[e.g.][]{2016ApJ...824..130G}, the velocity changes in S5 have been minimal, mostly staying within the margin of error through each epoch.\par
\section{Summary and Conclusion} \label{sec:conclusion}
From the VLT/UVES spectrum of the quasar SDSS J2357-0048, we have identified a BAL outflow with four subcomponents, S1, S2, S3, and S4, as well as a high-velocity \ion{C}{iv} mini-BAL, which we label as S5. This paper has presented the analysis of S1, based on the column densities of 14 ions shown in Table \ref{table:coldensity}. We have found estimates of $N_H$ and $U_H$ through photoionization analysis.\par
We used the abundance ratios between different energy states of \ion{O}{i} and \ion{Fe}{ii} to find the electron number density $n_e$ of S1, as shown in Fig.~\ref{fig:ratioplot}. We have also found other physical parameters of S1 such as distance from the quasar, mass flow rate, and kinetic luminosity, using Equations (\ref{eq:UH}), (\ref{eq:Mdot}), and (\ref{eq:Ekdot}). Based on the ratio between the kinetic luminosity $\dot{E}_k$ and the Eddington luminosity $L_{Edd}$ (see Table \ref{table:energetics}), the outflow's ability to contribute to AGN feedback has been found to be dependent on the model.\par
We also examined the time-variability of the S5 \ion{C}{iv} mini-BAL, based on the SDSS and UVES spectra (see Fig.~\ref{fig:bigBALvelocities}). The small absorption feature at $v=-7700\text{ km s}^{-1}$ became deeper between the 2000 and 2001 epochs, and nearly vanished in the 2005 epoch. The mini-BAL itself appears to have become gradually shallower over time.

\section*{Acknowledgements}

NA and DB acknowledge support from NSF grant AST 2106249, as well as NASA STScI grants GO 14777, 14242, 14054, 14176, and AR-15786. DB acknowledges support from the Virginia Space Grant Consortium Graduate Research Fellowship Program. We also thank the anonymous referee for their constructive comments that helped improve this paper.

\section*{Data Availability}

The normalized UVES spectrum of J2357-0048 is part of the SQUAD database made available by \citet{michael_murphy_2018_1463251} and described by \citet{Murphy2019}. The SDSS spectra used for the time variability analysis can be found in the SDSS archive \citep{2002AJ....123..485S,2011ApJS..193...29A,2012ApJS..203...21A,2017ApJS..233...25A}.



\bibliographystyle{mnras}
\bibliography{j2357} 

\begin{thebibliography}{}
\makeatletter
\relax
\def\mn@urlcharsother{\let\do\@makeother \do\$\do\&\do\#\do\^\do\_\do\%\do\~}
\def\mn@doi{\begingroup\mn@urlcharsother \@ifnextchar [ {\mn@doi@}
  {\mn@doi@[]}}
\def\mn@doi@[#1]#2{\def\@tempa{#1}\ifx\@tempa\@empty \href
  {http://dx.doi.org/#2} {doi:#2}\else \href {http://dx.doi.org/#2} {#1}\fi
  \endgroup}
\def\mn@eprint#1#2{\mn@eprint@#1:#2::\@nil}
\def\mn@eprint@arXiv#1{\href {http://arxiv.org/abs/#1} {{\tt arXiv:#1}}}
\def\mn@eprint@dblp#1{\href {http://dblp.uni-trier.de/rec/bibtex/#1.xml}
  {dblp:#1}}
\def\mn@eprint@#1:#2:#3:#4\@nil{\def\@tempa {#1}\def\@tempb {#2}\def\@tempc
  {#3}\ifx \@tempc \@empty \let \@tempc \@tempb \let \@tempb \@tempa \fi \ifx
  \@tempb \@empty \def\@tempb {arXiv}\fi \@ifundefined
  {mn@eprint@\@tempb}{\@tempb:\@tempc}{\expandafter \expandafter \csname
  mn@eprint@\@tempb\endcsname \expandafter{\@tempc}}}

\bibitem[\protect\citeauthoryear{{Ahn} et~al.,}{{Ahn}
  et~al.}{2012}]{2012ApJS..203...21A}
{Ahn} C.~P.,  et~al., 2012, \mn@doi [\apjs] {10.1088/0067-0049/203/2/21}, \href
  {https://ui.adsabs.harvard.edu/abs/2012ApJS..203...21A} {203, 21}

\bibitem[\protect\citeauthoryear{{Aihara} et~al.,}{{Aihara}
  et~al.}{2011}]{2011ApJS..193...29A}
{Aihara} H.,  et~al., 2011, \mn@doi [\apjs] {10.1088/0067-0049/193/2/29}, \href
  {https://ui.adsabs.harvard.edu/abs/2011ApJS..193...29A} {193, 29}

\bibitem[\protect\citeauthoryear{{Albareti} et~al.,}{{Albareti}
  et~al.}{2017}]{2017ApJS..233...25A}
{Albareti} F.~D.,  et~al., 2017, \mn@doi [\apjs] {10.3847/1538-4365/aa8992},
  \href {https://ui.adsabs.harvard.edu/abs/2017ApJS..233...25A} {233, 25}

\bibitem[\protect\citeauthoryear{Arav, Korista, de Kool, Junkkarinen  \&
  Begelman}{Arav et~al.}{1999a}]{Arav1999}
Arav N.,  Korista K.~T.,  de Kool M.,  Junkkarinen V.~T.,   Begelman M.~C.,
  1999a, \mn@doi [\apj] {10.1086/307073}, 516, 27

\bibitem[\protect\citeauthoryear{{Arav}, {Becker}, {Laurent-Muehleisen},
  {Gregg}, {White}, {Brotherton}  \& {de Kool}}{{Arav}
  et~al.}{1999b}]{1999ApJ...524..566A}
{Arav} N.,  {Becker} R.~H.,  {Laurent-Muehleisen} S.~A.,  {Gregg} M.~D.,
  {White} R.~L.,  {Brotherton} M.~S.,   {de Kool} M.,  1999b, \mn@doi [\apj]
  {10.1086/307841}, \href
  {https://ui.adsabs.harvard.edu/abs/1999ApJ...524..566A} {524, 566}

\bibitem[\protect\citeauthoryear{{Arav}, {Kaastra}, {Kriss}, {Korista}, {Gabel}
   \& {Proga}}{{Arav} et~al.}{2005}]{2005ApJ...620..665A}
{Arav} N.,  {Kaastra} J.,  {Kriss} G.~A.,  {Korista} K.~T.,  {Gabel} J.,
  {Proga} D.,  2005, \mn@doi [\apj] {10.1086/425560}, \href
  {https://ui.adsabs.harvard.edu/abs/2005ApJ...620..665A} {620, 665}

\bibitem[\protect\citeauthoryear{{Arav}, {Borguet}, {Chamberlain}, {Edmonds}
  \& {Danforth}}{{Arav} et~al.}{2013}]{2013MNRAS.436.3286A}
{Arav} N.,  {Borguet} B.,  {Chamberlain} C.,  {Edmonds} D.,   {Danforth} C.,
  2013, \mn@doi [\mnras] {10.1093/mnras/stt1812}, \href
  {https://ui.adsabs.harvard.edu/abs/2013MNRAS.436.3286A} {436, 3286}

\bibitem[\protect\citeauthoryear{{Arav} et~al.,}{{Arav}
  et~al.}{2015}]{2015A&A...577A..37A}
{Arav} N.,  et~al., 2015, \mn@doi [\aap] {10.1051/0004-6361/201425302}, \href
  {https://ui.adsabs.harvard.edu/abs/2015A&A...577A..37A} {577, A37}

\bibitem[\protect\citeauthoryear{{Arav}, {Liu}, {Xu}, {Stidham}, {Benn}  \&
  {Chamberlain}}{{Arav} et~al.}{2018}]{2018ApJ...857...60A}
{Arav} N.,  {Liu} G.,  {Xu} X.,  {Stidham} J.,  {Benn} C.,   {Chamberlain} C.,
  2018, \mn@doi [\apj] {10.3847/1538-4357/aab494}, \href
  {https://ui.adsabs.harvard.edu/abs/2018ApJ...857...60A} {857, 60}

\bibitem[\protect\citeauthoryear{{Arav}, {Xu}, {Miller}, {Kriss}  \&
  {Plesha}}{{Arav} et~al.}{2020}]{2020ApJS..247...37A}
{Arav} N.,  {Xu} X.,  {Miller} T.,  {Kriss} G.~A.,   {Plesha} R.,  2020,
  \mn@doi [\apjs] {10.3847/1538-4365/ab66af}, \href
  {https://ui.adsabs.harvard.edu/abs/2020ApJS..247...37A} {247, 37}

\bibitem[\protect\citeauthoryear{{Astropy Collaboration} et~al.,}{{Astropy
  Collaboration} et~al.}{2013}]{astropy:2013}
{Astropy Collaboration} et~al., 2013, \mn@doi [\aap]
  {10.1051/0004-6361/201322068}, \href
  {http://adsabs.harvard.edu/abs/2013A%26A...558A..33A} {558, A33}

\bibitem[\protect\citeauthoryear{{Astropy Collaboration} et~al.,}{{Astropy
  Collaboration} et~al.}{2018}]{astropy:2018}
{Astropy Collaboration} et~al., 2018, \mn@doi [\aj] {10.3847/1538-3881/aabc4f},
  \href {https://ui.adsabs.harvard.edu/abs/2018AJ....156..123A} {156, 123}

\bibitem[\protect\citeauthoryear{{Ballero}, {Matteucci}, {Ciotti}, {Calura}  \&
  {Padovani}}{{Ballero} et~al.}{2008}]{2008A&A...478..335B}
{Ballero} S.~K.,  {Matteucci} F.,  {Ciotti} L.,  {Calura} F.,   {Padovani} P.,
  2008, \mn@doi [\aap] {10.1051/0004-6361:20078663}, \href
  {https://ui.adsabs.harvard.edu/abs/2008A&A...478..335B} {478, 335}

\bibitem[\protect\citeauthoryear{{Barlow}}{{Barlow}}{2003}]{2003sppp.conf..250B}
{Barlow} R.,  2003, in {Lyons} L.,  {Mount} R.,   {Reitmeyer} R.,  eds,
  Statistical Problems in Particle Physics, Astrophysics, and Cosmology. p.~250
  (\mn@eprint {arXiv} {physics/0401042})

\bibitem[\protect\citeauthoryear{{Barlow}, {Hamann}  \& {Sargent}}{{Barlow}
  et~al.}{1997}]{1997ASPC..128...13B}
{Barlow} T.~A.,  {Hamann} F.,   {Sargent} W.~L.~W.,  1997, in {Arav} N.,
  {Shlosman} I.,   {Weymann} R.~J.,  eds,  Astronomical Society of the Pacific
  Conference Series Vol. 128, Mass Ejection from Active Galactic Nuclei. p.~13
  (\mn@eprint {arXiv} {astro-ph/9705048})

\bibitem[\protect\citeauthoryear{Bennett, Larson, Weiland  \& Hinshaw}{Bennett
  et~al.}{2014}]{Bennett_2014}
Bennett C.~L.,  Larson D.,  Weiland J.~L.,   Hinshaw G.,  2014, \mn@doi [\apj]
  {10.1088/0004-637x/794/2/135}, 794, 135

\bibitem[\protect\citeauthoryear{{Borguet}, {Edmonds}, {Arav}, {Dunn}  \&
  {Kriss}}{{Borguet} et~al.}{2012a}]{2012ApJ...751..107B}
{Borguet} B. C.~J.,  {Edmonds} D.,  {Arav} N.,  {Dunn} J.,   {Kriss} G.~A.,
  2012a, \mn@doi [\apj] {10.1088/0004-637X/751/2/107}, \href
  {https://ui.adsabs.harvard.edu/abs/2012ApJ...751..107B} {751, 107}

\bibitem[\protect\citeauthoryear{{Borguet}, {Edmonds}, {Arav}, {Benn}  \&
  {Chamberlain}}{{Borguet} et~al.}{2012b}]{2012ApJ...758...69B}
{Borguet} B. C.~J.,  {Edmonds} D.,  {Arav} N.,  {Benn} C.,   {Chamberlain} C.,
  2012b, \mn@doi [\apj] {10.1088/0004-637X/758/1/69}, \href
  {https://ui.adsabs.harvard.edu/abs/2012ApJ...758...69B} {758, 69}

\bibitem[\protect\citeauthoryear{{Byun}, {Arav}  \& {Hall}}{{Byun}
  et~al.}{2022}]{2022Byun}
{Byun} D.,  {Arav} N.,   {Hall} P.~B.,  2022, \mn@doi [\apj]
  {10.3847/1538-4357/ac503d}, \href
  {https://ui.adsabs.harvard.edu/abs/2022ApJ...927..176B} {927, 176}

\bibitem[\protect\citeauthoryear{{Chamberlain} \& {Arav}}{{Chamberlain} \&
  {Arav}}{2015}]{2015MNRAS.454..675C}
{Chamberlain} C.,  {Arav} N.,  2015, \mn@doi [\mnras] {10.1093/mnras/stv1979},
  \href {https://ui.adsabs.harvard.edu/abs/2015MNRAS.454..675C} {454, 675}

\bibitem[\protect\citeauthoryear{{Chamberlain}, {Arav}  \&
  {Benn}}{{Chamberlain} et~al.}{2015}]{2015MNRAS.450.1085C}
{Chamberlain} C.,  {Arav} N.,   {Benn} C.,  2015, \mn@doi [\mnras]
  {10.1093/mnras/stv572}, \href
  {https://ui.adsabs.harvard.edu/abs/2015MNRAS.450.1085C} {450, 1085}

\bibitem[\protect\citeauthoryear{{Chen}, {Hamann}, {Ma}  \& {Murphy}}{{Chen}
  et~al.}{2021}]{2021ApJ...907...84C}
{Chen} C.,  {Hamann} F.,  {Ma} B.,   {Murphy} M.,  2021, \mn@doi [\apj]
  {10.3847/1538-4357/abcec5}, \href
  {https://ui.adsabs.harvard.edu/abs/2021ApJ...907...84C} {907, 84}

\bibitem[\protect\citeauthoryear{{Coatman}, {Hewett}, {Banerji}, {Richards},
  {Hennawi}  \& {Prochaska}}{{Coatman} et~al.}{2017}]{2017MNRAS.465.2120C}
{Coatman} L.,  {Hewett} P.~C.,  {Banerji} M.,  {Richards} G.~T.,  {Hennawi}
  J.~F.,   {Prochaska} J.~X.,  2017, \mn@doi [\mnras] {10.1093/mnras/stw2797},
  \href {https://ui.adsabs.harvard.edu/abs/2017MNRAS.465.2120C} {465, 2120}

\bibitem[\protect\citeauthoryear{{Dai}, {Shankar}  \& {Sivakoff}}{{Dai}
  et~al.}{2008}]{2008ApJ...672..108D}
{Dai} X.,  {Shankar} F.,   {Sivakoff} G.~R.,  2008, \mn@doi [\apj]
  {10.1086/523688}, \href
  {https://ui.adsabs.harvard.edu/abs/2008ApJ...672..108D} {672, 108}

\bibitem[\protect\citeauthoryear{{Dere}, {Landi}, {Mason}, {Monsignori Fossi}
  \& {Young}}{{Dere} et~al.}{1997}]{1997A&AS..125..149D}
{Dere} K.~P.,  {Landi} E.,  {Mason} H.~E.,  {Monsignori Fossi} B.~C.,   {Young}
  P.~R.,  1997, \mn@doi [\aaps] {10.1051/aas:1997368}, \href
  {https://ui.adsabs.harvard.edu/abs/1997A&AS..125..149D} {125, 149}

\bibitem[\protect\citeauthoryear{Dere, Zanna, Young, Landi  \& Sutherland}{Dere
  et~al.}{2019}]{Dere_2019}
Dere K.~P.,  Zanna G.~D.,  Young P.~R.,  Landi E.,   Sutherland R.~S.,  2019,
  \mn@doi [\apjs] {10.3847/1538-4365/ab05cf}, 241, 22

\bibitem[\protect\citeauthoryear{{Edmonds} et~al.,}{{Edmonds}
  et~al.}{2011}]{2011ApJ...739....7E}
{Edmonds} D.,  et~al., 2011, \mn@doi [\apj] {10.1088/0004-637X/739/1/7}, \href
  {https://ui.adsabs.harvard.edu/abs/2011ApJ...739....7E} {739, 7}

\bibitem[\protect\citeauthoryear{{Ferland} et~al.,}{{Ferland}
  et~al.}{2017}]{2017RMxAA..53..385F}
{Ferland} G.~J.,  et~al., 2017, \rmxaa, \href
  {https://ui.adsabs.harvard.edu/abs/2017RMxAA..53..385F} {53, 385}

\bibitem[\protect\citeauthoryear{{Fitzpatrick}}{{Fitzpatrick}}{1999}]{1999PASP..111...63F}
{Fitzpatrick} E.~L.,  1999, \mn@doi [\pasp] {10.1086/316293}, \href
  {https://ui.adsabs.harvard.edu/abs/1999PASP..111...63F} {111, 63}

\bibitem[\protect\citeauthoryear{{Gabel} et~al.,}{{Gabel}
  et~al.}{2005}]{2005ApJ...631..741G}
{Gabel} J.~R.,  et~al., 2005, \mn@doi [\apj] {10.1086/432682}, \href
  {https://ui.adsabs.harvard.edu/abs/2005ApJ...631..741G} {631, 741}

\bibitem[\protect\citeauthoryear{{Grier} et~al.,}{{Grier}
  et~al.}{2016}]{2016ApJ...824..130G}
{Grier} C.~J.,  et~al., 2016, \mn@doi [\apj] {10.3847/0004-637X/824/2/130},
  \href {https://ui.adsabs.harvard.edu/abs/2016ApJ...824..130G} {824, 130}

\bibitem[\protect\citeauthoryear{{Hamann}, {Barlow}, {Chaffee}, {Foltz}  \&
  {Weymann}}{{Hamann} et~al.}{2001}]{2001ApJ...550..142H}
{Hamann} F.~W.,  {Barlow} T.~A.,  {Chaffee} F.~C.,  {Foltz} C.~B.,   {Weymann}
  R.~J.,  2001, \mn@doi [\apj] {10.1086/319733}, \href
  {https://ui.adsabs.harvard.edu/abs/2001ApJ...550..142H} {550, 142}

\bibitem[\protect\citeauthoryear{{He} et~al.,}{{He}
  et~al.}{2022}]{2022arXiv220206227H}
{He} Z.,  et~al., 2022, arXiv e-prints, \href
  {https://ui.adsabs.harvard.edu/abs/2022arXiv220206227H} {p. arXiv:2202.06227}

\bibitem[\protect\citeauthoryear{{Hewett} \& {Foltz}}{{Hewett} \&
  {Foltz}}{2003}]{2003AJ....125.1784H}
{Hewett} P.~C.,  {Foltz} C.~B.,  2003, \mn@doi [\aj] {10.1086/368392}, \href
  {https://ui.adsabs.harvard.edu/abs/2003AJ....125.1784H} {125, 1784}

\bibitem[\protect\citeauthoryear{{Hopkins} \& {Elvis}}{{Hopkins} \&
  {Elvis}}{2010}]{2010MNRAS.401....7H}
{Hopkins} P.~F.,  {Elvis} M.,  2010, \mn@doi [\mnras]
  {10.1111/j.1365-2966.2009.15643.x}, \href
  {https://ui.adsabs.harvard.edu/abs/2010MNRAS.401....7H} {401, 7}

\bibitem[\protect\citeauthoryear{{Knigge}, {Scaringi}, {Goad}  \&
  {Cottis}}{{Knigge} et~al.}{2008}]{2008MNRAS.386.1426K}
{Knigge} C.,  {Scaringi} S.,  {Goad} M.~R.,   {Cottis} C.~E.,  2008, \mn@doi
  [\mnras] {10.1111/j.1365-2966.2008.13081.x}, \href
  {https://ui.adsabs.harvard.edu/abs/2008MNRAS.386.1426K} {386, 1426}

\bibitem[\protect\citeauthoryear{{Leighly}, {Terndrup}, {Gallagher}, {Richards}
   \& {Dietrich}}{{Leighly} et~al.}{2018}]{2018ApJ...866....7L}
{Leighly} K.~M.,  {Terndrup} D.~M.,  {Gallagher} S.~C.,  {Richards} G.~T.,
  {Dietrich} M.,  2018, \mn@doi [\apj] {10.3847/1538-4357/aadee6}, \href
  {https://ui.adsabs.harvard.edu/abs/2018ApJ...866....7L} {866, 7}

\bibitem[\protect\citeauthoryear{{Miller}, {Arav}, {Xu}, {Kriss}  \&
  {Plesha}}{{Miller} et~al.}{2020a}]{2020ApJS..247...39M}
{Miller} T.~R.,  {Arav} N.,  {Xu} X.,  {Kriss} G.~A.,   {Plesha} R.~J.,  2020a,
  \mn@doi [\apjs] {10.3847/1538-4365/ab5967}, \href
  {https://ui.adsabs.harvard.edu/abs/2020ApJS..247...39M} {247, 39}

\bibitem[\protect\citeauthoryear{{Miller}, {Arav}, {Xu}, {Kriss}  \&
  {Plesha}}{{Miller} et~al.}{2020b}]{2020ApJS..247...41M}
{Miller} T.~R.,  {Arav} N.,  {Xu} X.,  {Kriss} G.~A.,   {Plesha} R.~J.,  2020b,
  \mn@doi [\apjs] {10.3847/1538-4365/ab5969}, \href
  {https://ui.adsabs.harvard.edu/abs/2020ApJS..247...41M} {247, 41}

\bibitem[\protect\citeauthoryear{{Miller}, {Arav}, {Xu}, {Kriss}  \&
  {Plesha}}{{Miller} et~al.}{2020c}]{2020ApJS..249...15M}
{Miller} T.~R.,  {Arav} N.,  {Xu} X.,  {Kriss} G.~A.,   {Plesha} R.~J.,  2020c,
  \mn@doi [\apjs] {10.3847/1538-4365/ab94b9}, \href
  {https://ui.adsabs.harvard.edu/abs/2020ApJS..249...15M} {249, 15}

\bibitem[\protect\citeauthoryear{{Miller}, {Arav}, {Xu}  \& {Kriss}}{{Miller}
  et~al.}{2020d}]{2020MNRAS.499.1522M}
{Miller} T.~R.,  {Arav} N.,  {Xu} X.,   {Kriss} G.~A.,  2020d, \mn@doi [\mnras]
  {10.1093/mnras/staa2981}, \href
  {https://ui.adsabs.harvard.edu/abs/2020MNRAS.499.1522M} {499, 1522}

\bibitem[\protect\citeauthoryear{{Moe}, {Arav}, {Bautista}  \& {Korista}}{{Moe}
  et~al.}{2009}]{2009ApJ...706..525M}
{Moe} M.,  {Arav} N.,  {Bautista} M.~A.,   {Korista} K.~T.,  2009, \mn@doi
  [\apj] {10.1088/0004-637X/706/1/525}, \href
  {https://ui.adsabs.harvard.edu/abs/2009ApJ...706..525M} {706, 525}

\bibitem[\protect\citeauthoryear{Murphy}{Murphy}{2018}]{michael_murphy_2018_1463251}
Murphy M.,  2018, {MTMurphy77/UVES\_SQUAD\_DR1: First data release of the UVES
  Spectral Quasar Absorption Database (SQUAD)},
  \mn@doi{10.5281/zenodo.1463251}, \url
  {https://doi.org/10.5281/zenodo.1463251}

\bibitem[\protect\citeauthoryear{Murphy, Kacprzak, Savorgnan  \&
  Carswell}{Murphy et~al.}{2019}]{Murphy2019}
Murphy M.~T.,  Kacprzak G.~G.,  Savorgnan G.~A.,   Carswell R.~F.,  2019,
  \mn@doi [\mnras] {10.1093/mnras/sty2834}, 482, 3458

\bibitem[\protect\citeauthoryear{{Osterbrock} \& {Ferland}}{{Osterbrock} \&
  {Ferland}}{2006}]{2006agna.book.....O}
{Osterbrock} D.~E.,  {Ferland} G.~J.,  2006, {Astrophysics of gaseous nebulae
  and active galactic nuclei}

\bibitem[\protect\citeauthoryear{Savage \& Sembach}{Savage \&
  Sembach}{1991}]{Savage1991}
Savage B.~D.,  Sembach K.~R.,  1991, \mn@doi [\apj] {10.1086/170498}, 379, 245

\bibitem[\protect\citeauthoryear{{Scannapieco} \& {Oh}}{{Scannapieco} \&
  {Oh}}{2004}]{2004ApJ...608...62S}
{Scannapieco} E.,  {Oh} S.~P.,  2004, \mn@doi [\apj] {10.1086/386542}, \href
  {https://ui.adsabs.harvard.edu/abs/2004ApJ...608...62S} {608, 62}

\bibitem[\protect\citeauthoryear{{Schlafly} \& {Finkbeiner}}{{Schlafly} \&
  {Finkbeiner}}{2011}]{2011ApJ...737..103S}
{Schlafly} E.~F.,  {Finkbeiner} D.~P.,  2011, \mn@doi [\apj]
  {10.1088/0004-637X/737/2/103}, \href
  {https://ui.adsabs.harvard.edu/abs/2011ApJ...737..103S} {737, 103}

\bibitem[\protect\citeauthoryear{{Silk} \& {Rees}}{{Silk} \&
  {Rees}}{1998}]{1998A&A...331L...1S}
{Silk} J.,  {Rees} M.~J.,  1998, \aap, \href
  {https://ui.adsabs.harvard.edu/abs/1998A&A...331L...1S} {331, L1}

\bibitem[\protect\citeauthoryear{{Stoughton} et~al.,}{{Stoughton}
  et~al.}{2002}]{2002AJ....123..485S}
{Stoughton} C.,  et~al., 2002, \mn@doi [\aj] {10.1086/324741}, \href
  {https://ui.adsabs.harvard.edu/abs/2002AJ....123..485S} {123, 485}

\bibitem[\protect\citeauthoryear{{Vayner} et~al.,}{{Vayner}
  et~al.}{2021}]{2021ApJ...919..122V}
{Vayner} A.,  et~al., 2021, \mn@doi [\apj] {10.3847/1538-4357/ac0f56}, \href
  {https://ui.adsabs.harvard.edu/abs/2021ApJ...919..122V} {919, 122}

\bibitem[\protect\citeauthoryear{{Vestergaard} \& {Peterson}}{{Vestergaard} \&
  {Peterson}}{2006}]{2006ApJ...641..689V}
{Vestergaard} M.,  {Peterson} B.~M.,  2006, \mn@doi [\apj] {10.1086/500572},
  \href {https://ui.adsabs.harvard.edu/abs/2006ApJ...641..689V} {641, 689}

\bibitem[\protect\citeauthoryear{{Xu}, {Arav}, {Miller}  \& {Benn}}{{Xu}
  et~al.}{2018}]{2018ApJ...858...39X}
{Xu} X.,  {Arav} N.,  {Miller} T.,   {Benn} C.,  2018, \mn@doi [\apj]
  {10.3847/1538-4357/aab7ea}, \href
  {https://ui.adsabs.harvard.edu/abs/2018ApJ...858...39X} {858, 39}

\bibitem[\protect\citeauthoryear{{Xu}, {Arav}, {Miller}  \& {Benn}}{{Xu}
  et~al.}{2019}]{2019ApJ...876..105X}
{Xu} X.,  {Arav} N.,  {Miller} T.,   {Benn} C.,  2019, \mn@doi [\apj]
  {10.3847/1538-4357/ab164e}, \href
  {https://ui.adsabs.harvard.edu/abs/2019ApJ...876..105X} {876, 105}

\bibitem[\protect\citeauthoryear{{Xu}, {Arav}, {Miller}, {Kriss}  \&
  {Plesha}}{{Xu} et~al.}{2020a}]{2020ApJS..247...38X}
{Xu} X.,  {Arav} N.,  {Miller} T.,  {Kriss} G.~A.,   {Plesha} R.,  2020a,
  \mn@doi [\apjs] {10.3847/1538-4365/ab596a}, \href
  {https://ui.adsabs.harvard.edu/abs/2020ApJS..247...38X} {247, 38}

\bibitem[\protect\citeauthoryear{{Xu}, {Arav}, {Miller}, {Kriss}  \&
  {Plesha}}{{Xu} et~al.}{2020b}]{2020ApJS..247...40X}
{Xu} X.,  {Arav} N.,  {Miller} T.,  {Kriss} G.~A.,   {Plesha} R.,  2020b,
  \mn@doi [\apjs] {10.3847/1538-4365/ab4bcb}, \href
  {https://ui.adsabs.harvard.edu/abs/2020ApJS..247...40X} {247, 40}

\bibitem[\protect\citeauthoryear{{Xu}, {Arav}, {Miller}, {Kriss}  \&
  {Plesha}}{{Xu} et~al.}{2020c}]{2020ApJS..247...42X}
{Xu} X.,  {Arav} N.,  {Miller} T.,  {Kriss} G.~A.,   {Plesha} R.,  2020c,
  \mn@doi [\apjs] {10.3847/1538-4365/ab5f68}, \href
  {https://ui.adsabs.harvard.edu/abs/2020ApJS..247...42X} {247, 42}

\bibitem[\protect\citeauthoryear{{Yuan}, {Yoon}, {Li}, {Gan}, {Ho}  \&
  {Guo}}{{Yuan} et~al.}{2018}]{2018ApJ...857..121Y}
{Yuan} F.,  {Yoon} D.,  {Li} Y.-P.,  {Gan} Z.-M.,  {Ho} L.~C.,   {Guo} F.,
  2018, \mn@doi [\apj] {10.3847/1538-4357/aab8f8}, \href
  {https://ui.adsabs.harvard.edu/abs/2018ApJ...857..121Y} {857, 121}

\bibitem[\protect\citeauthoryear{{Zafar}, {Popping}  \& {P{\'e}roux}}{{Zafar}
  et~al.}{2013}]{2013A&A...556A.140Z}
{Zafar} T.,  {Popping} A.,   {P{\'e}roux} C.,  2013, \mn@doi [\aap]
  {10.1051/0004-6361/201321153}, \href
  {https://ui.adsabs.harvard.edu/abs/2013A&A...556A.140Z} {556, A140}

\bibitem[\protect\citeauthoryear{{de Kool}, {Arav}, {Becker}, {Gregg}, {White},
  {Laurent-Muehleisen}, {Price}  \& {Korista}}{{de Kool}
  et~al.}{2001}]{2001ApJ...548..609D}
{de Kool} M.,  {Arav} N.,  {Becker} R.~H.,  {Gregg} M.~D.,  {White} R.~L.,
  {Laurent-Muehleisen} S.~A.,  {Price} T.,   {Korista} K.~T.,  2001, \mn@doi
  [\apj] {10.1086/318996}, \href
  {https://ui.adsabs.harvard.edu/abs/2001ApJ...548..609D} {548, 609}

\bibitem[\protect\citeauthoryear{{de Kool}, {Becker}, {Gregg}, {White}  \&
  {Arav}}{{de Kool} et~al.}{2002a}]{2002ApJ...567...58D}
{de Kool} M.,  {Becker} R.~H.,  {Gregg} M.~D.,  {White} R.~L.,   {Arav} N.,
  2002a, \mn@doi [\apj] {10.1086/338490}, \href
  {https://ui.adsabs.harvard.edu/abs/2002ApJ...567...58D} {567, 58}

\bibitem[\protect\citeauthoryear{de Kool, Korista  \& Arav}{de~Kool
  et~al.}{2002b}]{DeKool2002}
de Kool M.,  Korista K.~T.,   Arav N.,  2002b, \mn@doi [\apj] {10.1086/343107},
  580, 54

\makeatother
\end{thebibliography}





\bsp	
\label{lastpage}
\end{document}